\documentclass[10pt]{book}
\usepackage{array}
\usepackage{natbib,graphicx,amssymb,bm,setspace,dcolumn,amsmath,amsthm} % PJN
\usepackage{chngcntr} % PJN 
\counterwithout{figure}{chapter} % PJN

\newtheorem{thm}{Theorem}

\newtheoremstyle{dotless}{}{}{\itshape}{}{\bfseries}{}{ }{}

\theoremstyle{dotless}

\newtheorem*{cor}{Corollary}
\let\leq=\leqslant
\let\geq=\geqslant

\newcommand{\bit}{\begin{itemize}}
\newcommand{\eit}{\end{itemize}}
\newcommand{\beqnn}{\begin{eqnarray*}}
\newcommand{\eeqnn}{\end{eqnarray*}}
\newcommand{\beqn}{\begin{eqnarray}}
\newcommand{\eeqn}{\end{eqnarray}}
\newcommand{\e}{{\rm e}}

\begin{document}

%%%%%%%%%%%%%%%%%%%%%%%%%%%%%%%%%%%%%%%%%%%%%%%%%%%%%%%%%%%%%%%%%%%%%%%%%%%%%%%%%%%%%%%%%%%%%%%%%%%%%%%%%%%%%%%%%%%%%%%%%%%%
%%%%%%%%%%%%%%%%%%%%%%%%%%%%%%%%%%%%%%%%%%%%%%%%%%%%%%%%%%%%%%%%%%%%%%%%%%%%%%%%%%%%%%%%%%%%%%%%%%%%%%%%%%%%%%%%%%%%%%%%%%%%

\renewcommand{\baselinestretch}{1.2}

\markright{
%\hbox{\footnotesize\rm Statistica Sinica (2014): Preprint}\hfill
\hbox{\footnotesize\rm }\hfill
}

\markboth{\hfill{\footnotesize\rm PAUL NORTHROP AND NICOLAS ATTALIDES} \hfill}
{\hfill {\footnotesize\rm REFRENCE BAYESIAN EXTREME VALUE ANALYSES} \hfill}

\renewcommand{\thefootnote}{}
$\ $\par

%%%%%%%%%%%%%%%%%%%%%%%%%%%%%%%%%%%%%%%%%%%%%%%%%%%%%%%%%%%%%%%%%%%%%%%%%%%%%%%%%%%%%%%%%%%%%%%%%%%%%%%%%%%%%%%%%%%%%%%%%%%%

\fontsize{10.95}{14pt plus.8pt minus .6pt}\selectfont
\vspace{0.8pc}
\centerline{\large\bf POSTERIOR PROPRIETY IN BAYESIAN EXTREME VALUE}
\vspace{2pt}
\centerline{\large\bf  ANALYSES USING REFERENCE PRIORS}
\vspace{.4cm}
\centerline{Paul J. Northrop\footnote{
%\hspace*{-0.8cm}
{\it Address for correspondence}: Paul Northrop, Department of Statistical Science, University College London, Gower Street, London WC1E 6BT, UK. 
E-mail: p.northrop@ucl.ac.uk} and Nicolas Attalides}
\vspace{.4cm}
\centerline{\it University College London}
\vspace{.55cm}
\fontsize{9}{11.5pt plus.8pt minus .6pt}\selectfont

%%%%%%%%%%%%%%%%%%%%%%%%%%%%%%%%%%%%%%%%%%%%%%%%%%%%%%%%%%%%%%%%%%%%%%%%%%%%%%%%%%%%%%%%%%%%%%%%%%%%%%%%%%%%%%%%%%%%%%%%%%%%

\begin{quotation}
\noindent {\it Abstract:}
The Generalized Pareto (GP) and Generalized extreme value (GEV) distributions play an important role in extreme value analyses, as models for threshold excesses and block maxima respectively.  
For each of these distributions we consider Bayesian inference using ``reference'' prior distributions
(in the general sense of priors constructed using formal rules)
for the model parameters, specifically a Jeffreys prior, the maximal data information (MDI) prior and independent uniform priors on separate model parameters.
We investigate the important issue of whether these improper priors lead to proper posterior distributions.
We show that, in the GP and GEV cases, the MDI prior, unless modified, never yields a proper posterior and that in the GEV case this also applies to the Jeffreys prior.
We also show that a sample size of three (four) is sufficient for independent uniform priors to yield a proper posterior distribution in the GP (GEV) case. 
\par

\vspace{9pt}
\par
\end{quotation}\par

%%%%%%%%%%%%%%%%%%%%%%%%%%%%%%%%%%%%%%%%%%%%%%%%%%%%%%%%%%%%%%%%%%%%%%%%%%%%%%%%%%%%%%%%%%%%%%%%%%%%%%%%%%%%%%%%%%%%%%%%%%%%

%\setcounter{figure}{1}

\fontsize{10.95}{14pt plus.8pt minus .6pt}\selectfont

\setcounter{chapter}{1}
\setcounter{equation}{0} %-1
\noindent {\bf 1. Introduction}

Extreme value theory provides an asymptotic justification for particular families of models for extreme data.
Let $X_1, X_2, \ldots X_N$ be a sequence of independent and identically distributed random variables.
Let $u_N$ be a threshold, increasing with $N$.  
\cite{Pickands1975} showed that if there is a non-degenerate limiting
distribution for appropriately linearly rescaled excesses of $u_N$ then this limit is a Generalized Pareto (GP) distribution.
In practice, a suitably high threshold $u$ is chosen empirically.
Given that there is an exceedance of $u$, the excess $Z=X-u$ is modelled by a GP($\sigma_u,\xi$) distribution, with threshold-dependent scale parameter $\sigma_u$, shape parameter $\xi$ and distribution function
\begin{equation}
F_{GP}(z)= 
\begin{cases} 
1-\left( 1+\xi z /\sigma_u\right)_+^{-1/\xi}, & \xi \neq 0, \\
1-\exp(-z/\sigma_u), &  \xi = 0,
\end{cases} \label{eqn:GP}
\end{equation}
where $z>0$, $z_+=\max(z,0)$, $\sigma_u>0$ and $\xi \in \mathbb{R}$.
The use of the generalized extreme value (GEV)
distribution \citep{Jenkinson1955}, with distribution function
\begin{equation}
F_{GEV}(y)= 
\begin{cases} 
\exp\left\{ -\left[ 1+\xi (y-\mu) /\sigma \right]_+^{-1/\xi} \right\}, & \xi \neq 0, \\
\exp\left\{ - \exp [ -  (y-\mu) /\sigma ] \right\}, &  \xi = 0,
\end{cases} \label{eqn:GEV}
\end{equation}
where $\sigma>0$ and $\mu, \xi \in \mathbb{R}$,
as a model for block maxima is motivated by considering the
behaviour of $Y=\max\{X_1, \ldots, X_b\}$ as $b \rightarrow \infty$
\citep{FT1928, LLR1983}. 

Commonly-used frequentist methods of inference for extreme value distributions are maximum likelihood estimation (MLE) and probability-weighted moments (PWM).
However, conditions on $\xi$ are required for the asymptotic theory on which inferences are based to apply: $\xi > -1/2$ for MLE \citep{Smith1984,Smith1985} and $\xi < 1/2$ for PWM \citep{Hosking1985,Hosking1987}.
Alternatively, a Bayesian approach \citep{Coles2001,CP1996,ST2004} can avoid conditions on the value of $\xi$ and performs predictive inference about future observations naturally and conveniently using Markov chain Monte Carlo (MCMC) output.
A distinction can be made between {\it subjective} analyses, in which the prior distribution supplies information from an expert \citep{Coles1996} or more general experience of the quantity under study \citep{Martins2000,Martins2001}, and so-called {\it objective} analyses \citep{Berger2006}.  In the latter, a prior is constructed using a formal rule, for use when no subjective information is to be incorporated into the analysis.  There is disagreement about appropriate terminology for such priors: we follow \cite{Kass1996} in using the term {\it reference prior}.

Many such formal rules have been proposed: \cite{Kass1996} provides a comprehensive review.  In this paper we consider three priors that have been used in extreme value analyses:  the Jeffreys prior \citep{EugeniaCastellanos2007,Beirlant2004}, the maximal data information (MDI) prior \citep{Beirlant2004}, and the uniform prior \citep{Pickands1994}.  
These priors are {\it improper}, that is, they do not integrate to a finite number and therefore do not correspond to a proper probability distribution.
An improper prior can lead to an improper posterior, which is clearly undesirable.
There is no general theory providing simple conditions under which an improper prior yields a proper posterior for a particular model, so this must be investigated case-by-case.
\cite{EugeniaCastellanos2007} establish that Jeffreys prior for the GP distribution always yields a proper posterior, but no such results exist for the other improper priors we consider.
It is important that posterior propriety is established because impropriety may not create obvious numerical problems, for example, MCMC output may appear perfectly reasonable
\citep{Hobert1996}.

One way to ensure posterior propriety is to use a diffuse proper prior, such as a normal prior with a large variance \citep{CT2005,Smith2005} or by truncating an improper prior \citep{SG2000}.  
For example, \citet[chapter 9]{Coles2001} uses a GEV($\mu,\sigma,\xi$) model for annual maximum sea-levels, placing independent normal priors on $\mu$, $\log \sigma$ and $\xi$ with respective variances $10^4, 10^4$ and $100$.
However, one needs to check that the posterior is not sensitive to the choice of proper prior and, as \cite{Bayarri2004} note
``\ldots these posteriors will essentially be meaningless if
the limiting improper objective prior would have resulted
in an improper posterior distribution.''
Therefore, independent uniform priors on separate model parameters are of interest in their own right and as the limiting case of independent diffuse normal priors.

In section 2 we give the general form of the three priors we consider in this paper.
In section 3 we investigate whether or not these priors yield a proper posterior distribution given a random sample $\bm{z}=(z_1, \ldots, z_m)$
from the GP distribution,
and, in cases where propriety is possible, we derive sufficient conditions for this to occur.
We repeat this for a random sample $\bm{y}=(y_1, \ldots, y_n)$ from a GEV distribution in section 4.
In section 5 we discuss some implications of these results and possible extensions.
Proofs of results are presented in the appendix.
\par

\pagebreak

%%%%%%%%%%%%%%%%%%%%%%%%%%%%%%%%%%%%%%%%%%%%%%%%%%%%%%%%%%%%%%%%%%%%%%%%%%%%%%%%%%%%%%%%%%%%%%%%%%%%%%%%%%%%%%%%%%%%%%%%%%%%

\setcounter{chapter}{2}
\setcounter{equation}{0} %-1
\noindent {\bf 2. Reference priors for extreme value distributions}
\label{sec:priors}

Let $Y$ is a random variable with density function $f(Y~|~\phi)$, indexed by a parameter vector $\phi$, and define the Fisher information matrix 
$I(\phi)$ by $I(\phi)_{ij}={\rm E}\left[-\partial^2 \ln f(Y~|~\phi) / \partial \phi_i \partial \phi_j \right]$.

{\it Uniform priors.}  
Priors that are flat, i.e. equal to a positive constant, suffer from the problem that they are not automatically invariant to reparameterisation: for example, if we give $\log \sigma$ a uniform distributon then $\sigma$ is not uniform.
Thus, it matters which particular parameterization is used to define the prior.

{\it Jeffreys priors.}
Jeffreys' ``general rule'' \citep{Jeffreys1961} is
\begin{equation}
\pi_J(\phi) \propto \det (I(\phi))^{1/2}. \label{eqn:Jeff}
\end{equation}
An attractive property of this rule is that it produces a prior that is invariant to reparameterization.
Jeffreys suggested a modification of this rule for use in location-scale problems.
We will follow this modification, which is summarised on page 1345 of \cite{Kass1996}.
If there is no location parameter then (\ref{eqn:Jeff}) is used.
If there is a location parameter $\mu$, say, then $\phi=(\mu,\theta)$ and
\begin{equation}
\pi_J(\mu,\theta)\propto \det (I(\theta))^{1/2}, \label{eqn:Jeff_loc}
\end{equation}
where $I(\theta)$ is calculated holding $\mu$ fixed.
In the current context the GP distribution does not have a location parameter whereas the GEV distribution does.

{\it MDI prior.}
The MDI prior \citep{Zellner1971} is defined as
\begin{equation}
\pi_M(\phi) \propto \exp \left\{{\rm E}[\log f(Y~|~\phi)] \right\}. \label{eqn:MDI_prior}
\end{equation}
This is the prior for which the increase in average information, provided by the data via the likelihood function, is maximised.
For further information see \cite{Zellner1998}.
\par

%%%%%%%%%%%%%%%%%%%%%%%%%%%%%%%%%%%%%%%%%%%%%%%%%%%%%%%%%%%%%%%%%%%%%%%%%%%%%%%%%%%%%%%%%%%%%%%%%%%%%%%%%%%%%%%%%%%%%%%%%%%%

\setcounter{chapter}{3}
\setcounter{equation}{0} %-1
\noindent {\bf 3. Generalized Pareto (GP) distribution}
\label{sec:GP}

Without loss of generality we take the $m$ threshold excesses to be ordered: $z_1 < \cdots < z_m$.
For simplicity we denote the GP scale parameter by $\sigma$ rather than $\sigma_u$.
We consider a class of priors of the form $\pi(\sigma,\xi) \propto \pi(\xi)/\sigma, \sigma > 0, \xi \in \mathbb{R}$,
where $\pi(\xi)$ is a function depending only on $\xi$,
that is, {\it a priori} $\sigma$ and $\xi$ are independent and $\log \sigma$ has an improper uniform prior over the real line.

The posterior is given by
\begin{equation}
\pi_G(\sigma, \xi~|~\bm{z}) = C_m^{-1} \pi(\xi) \, \sigma^{-(m+1)} \prod_{i=1}^m 
\left( 1+\xi z_i / \sigma \right)^{-(1+1/\xi)}, \quad \sigma>0, \, \xi > -\sigma/z_m, \nonumber
\end{equation}
where
\begin{equation}
C_m = \int_{-\infty}^{\infty} \int_{\max(0,-\xi z_m)}^{\infty} 
\pi(\xi) \, \sigma^{-(m+1)} \prod_{i=1}^m \left( 1+\xi z_i / \sigma \right)^{-(1+1/\xi)}
{\rm ~d}\sigma {\rm ~d}\xi \label{eqn:const_GP}
\end{equation}
and the inequality $\xi > -\sigma/z_m$ comes from the constraints $1+\xi z_i /\sigma>0, i=1, \ldots, m$ in the likelihood.
\par

\pagebreak

\noindent {\bf 3.1 Prior densities}

Using (\ref{eqn:Jeff}) with $\phi=(\sigma,\xi)$ gives the Jeffreys prior 
\[ \pi_{J,GP}(\sigma,\xi) \propto \frac{1}{\sigma (1+\xi) (1 + 2 \xi)^{1/2}}, \quad \sigma > 0, \,\xi > -1/2. \]
\cite{EugeniaCastellanos2007} show that a proper posterior density results for $m \geq 1$.
% Smith (1984), BGST2004 page 447.

Using (\ref{eqn:MDI_prior}) gives the MDI prior
\begin{equation}
\pi_{M,GP}(\sigma,\xi) \propto \frac{1}{\sigma} \,{\rm e}^{-(\xi+1)} \propto \frac{1}{\sigma} \,{\rm e}^{-\xi} \quad \sigma > 0, \, \xi \in \mathbb{R}.  \label{MDI:GP}
\end{equation}
\citet[page 447]{Beirlant2004} use this prior but they do not investigate the propriety of the posterior.   

Placing independent uniform priors on $\log \sigma$ and $\xi$ gives the prior 
\begin{equation}
\pi_{U, GP}(\sigma,\xi) \propto \frac{1}{\sigma}, \qquad \sigma>0, \, \xi \in \mathbb{R}, \label{eqn:unif_GP}.
\end{equation}
%which is improper for both $\sigma$ and $\xi$.
This prior was proposed by \cite{Pickands1994}.  

\begin{figure}[h]
\centering
\includegraphics[width=0.75\textwidth, angle=0]{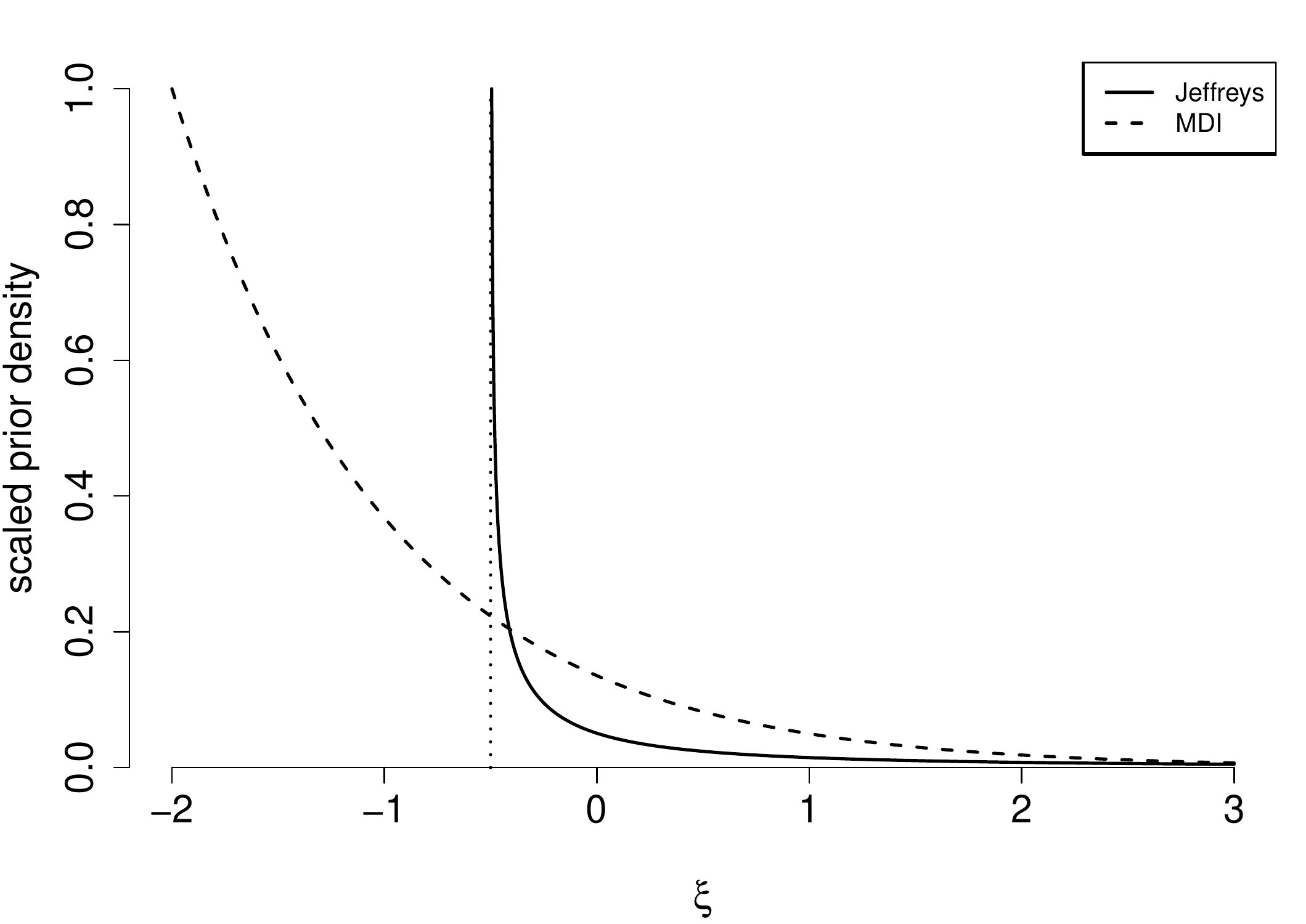}
\caption{\label{fig:GP_priors} Scaled Jeffreys and MDI GP prior densities against $\xi$.}
\end{figure}

Figure \ref{fig:GP_priors} shows the Jeffreys and MDI priors for GP parameters as a functions of $\xi$.  The MDI prior increases without limit as $\xi \rightarrow -\infty$.

\noindent {\bf 3.2 Results}

\begin{thm}\label{th:general_GP}
A sufficient condition for the prior $\pi(\sigma,\xi) \propto \pi(\xi) /\sigma, \sigma>0, \xi \in \mathbb{R}$ to yield a proper posterior density function is that $\pi(\xi)$ is (proportional to) a proper density function.
\end{thm}
The MDI prior (\ref{MDI:GP}) does not satisfy the condition in theorem \ref{th:general_GP} because $\exp\{-(\xi+1)\}$ is not a proper density function on $\xi \in \mathbb{R}$.

\begin{thm}\label{th:MDI_GP}
There is no sample size for which the MDI prior (\ref{MDI:GP}) yields a proper posterior density function.
\end{thm}

The problem with the MDI prior is due to its behaviour for negative $\xi$ so a simple solution is to place a lower bound on $\xi$ {\it a priori}.
This approach is common in extreme value analyses, for example, \cite{Martins2001} constrain $\xi$ to $(-1/2,1/2)$ {\it a priori}.
We suggest 
\begin{equation}
\pi'_{M,GP}(\sigma,\xi)=\frac{1}{\sigma}\e^{-(\xi+1)}, \, \xi \geq -1, \label{eqn:mod_GP}
\end{equation}
that is, a (proper) unit exponential prior on $\xi+1$.
Any finite lower bound on $\xi$ ensures propriety of the posterior but $\xi=-1$, for which the GP distribution reduces to a uniform distribution on $(0,\sigma)$, seems less arbitrary than other choices as it corresponds to a change in the behaviour of the GP density.  
For $\xi>-1$, the GP density $f_{GP}(z)$ decreases in $z$, which is what one anticipates when conducting an extreme value analysis to make inferences about future large, rare values.
For $\xi < -1$, $f_{GP}(z)$ increases without limit as it approaches its mode at the upper end point $-\sigma/\xi$, behaviour that is not expected in such analyses.

\begin{cor}\label{th:mod_MDI_GP}
{\bf to theorem \ref{th:general_GP}.} The truncated MDI prior (\ref{eqn:mod_GP}) yields a proper posterior density function for $m \geq 1$.
\end{cor}

% Comment re more than one observation is necessary because one observation is not sufficient to learn about both $\sigma$ and $\xi$?

\begin{thm}\label{th:F_GP}
A sufficient condition for the uniform prior (\ref{eqn:unif_GP}) to yield a proper posterior density function is that $m \geq 3$.
\end{thm}

\par

%%%%%%%%%%%%%%%%%%%%%%%%%%%%%%%%%%%%%%%%%%%%%%%%%%%%%%%%%%%%%%%%%%%%%%%%%%%%%%%%%%%%%%%%%%%%%%%%%%%%%%%%%%%%%%%%%%%%%%%%%%%%

\setcounter{chapter}{4}
\setcounter{equation}{0} %-1
\noindent {\bf 4. Generalized extreme value (GEV) distribution}
\label{sec:GEV}

Without loss of generality we take the $n$ block maxima to be ordered: $y_1 < \cdots < y_n$.
We consider a class of priors of the form $\pi(\mu,\sigma,\xi) \propto \pi(\xi)/\sigma, \sigma >0, \mu, \xi \in \mathbb{R}$ that is, {\it a priori} $\mu$, $\sigma$ and $\xi$ are independent and $\mu$ and $\log \sigma$ have improper uniform priors over the real line.

Based on a random sample $y_1, \ldots, y_n$ the posterior density for $(\mu,\sigma,\xi)$ is proportional to
\begin{equation}
\sigma^{-(n+1)} \pi(\xi)
\exp\left\{- \sum_{i=1}^n  z_i^{-1/\xi} \right\}
\prod_{i=1}^n z_i^{-(1+1/\xi)},
\label{eqn:GEVpost}
\end{equation}
where $z_i=1+\xi(y_i-\mu)/\sigma$ and $\sigma>0$.
If $\xi>0$ then $\mu-\sigma/\xi < y_1$ and
if $\xi < 0$ then $\mu-\sigma/\xi > y_n$.

\noindent {\bf 4.1 Prior densities}

\citet[page 63]{KN2000} give the Fisher information matrix for the GEV distribution (\ref{eqn:GEV}).
Using (\ref{eqn:Jeff_loc}) with $\phi=(\mu,\sigma,\xi)$ gives the Jeffreys prior 
\beqn
\pi_{J,GEV}(\mu,\sigma,\xi) &=& \frac{1}{\sigma \xi^2}
\Bigg\{ 
\left[ 1-2\Gamma(2+\xi) + p \right] 
\left[ \frac{\pi^2}{6}+\left(1-\gamma+\frac{1}{\xi}\right)^2 -\frac{2q}{\xi}+\frac{p}{\xi^2} \right] \nonumber \\ 
&& \hspace*{-2.2cm} - \left[ 1-\gamma+\frac{1}{\xi}-\frac{1}{\xi} \Gamma(2+\xi) -q + \frac{p}{\xi} \right]^2
\Bigg\}^{1/2}\!\!\!\!,
\,\, \mu \in \mathbb{R},\sigma>0, \,\xi > -1/2, \label{eqn:J_GEV}
\eeqn
where $p=(1+\xi)^2 \, \Gamma(1+2\xi)$, 
$q=\Gamma(2+\xi) \left\{ \psi(1+\xi)+(1+\xi)/\xi\right\}$,
$\psi(r)=\partial \log \Gamma(r) / \partial r$
and $\gamma \approx 0.57722$ is Euler's constant.
\cite{van2004} give an alternative form for the Jeffreys prior, based on (\ref{eqn:Jeff}).

\begin{figure}[h]
\centering
\includegraphics[width=0.75\textwidth, angle=0]{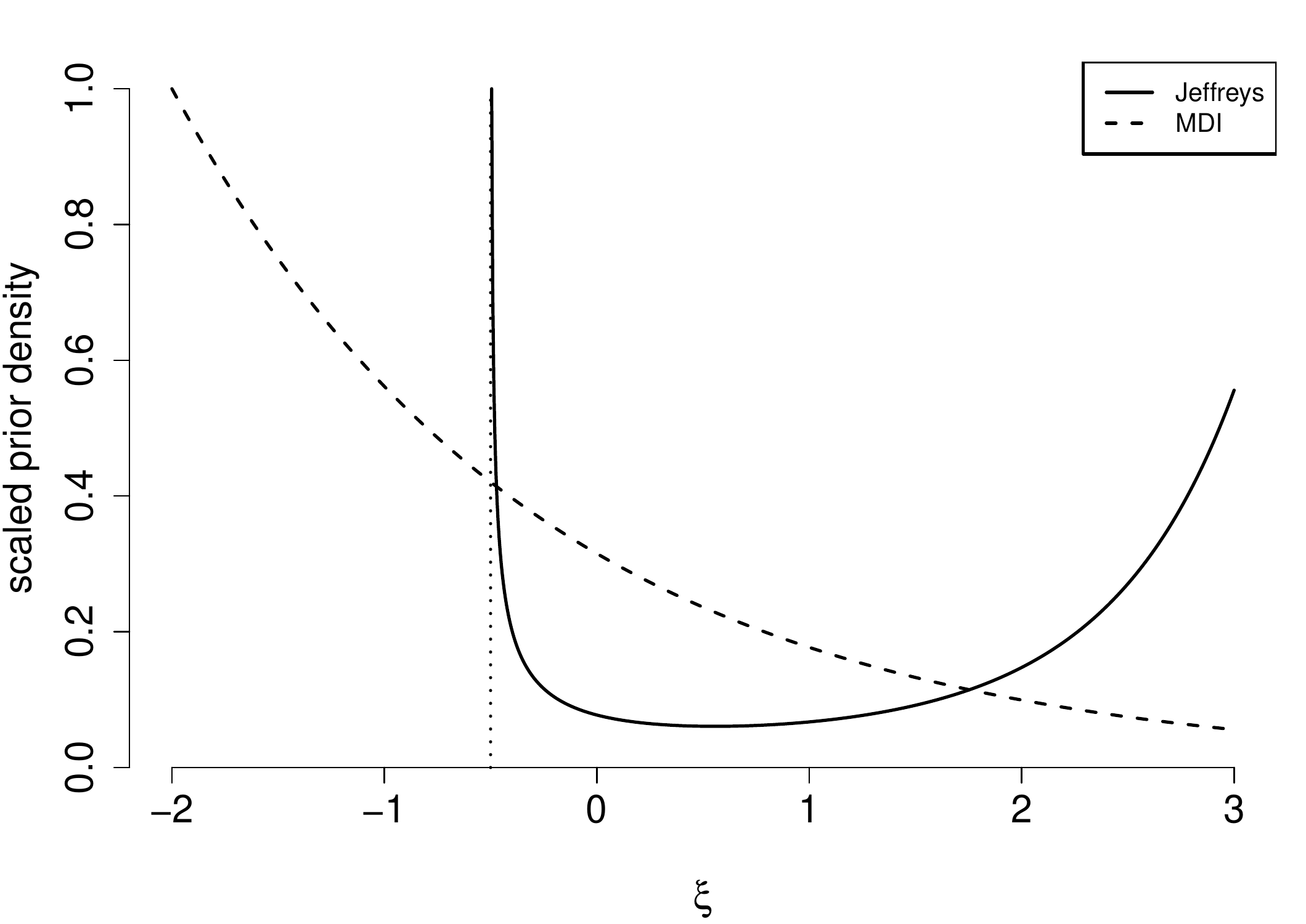}
\caption{\label{fig:GEV_priors} Scaled Jeffreys and MDI GEV prior densities against $\xi$.}
\end{figure}

\citet[page 435]{Beirlant2004} give the form of the MDI prior:
\begin{equation}
\pi_{M,GEV}(\mu,\sigma,\xi) = \frac{1}{\sigma} \,{\rm e}^{-\gamma(\xi+1+1/\gamma)} \propto \frac{1}{\sigma} \,{\rm e}^{-\gamma(1+\xi)}, \quad \sigma > 0, \, \mu, \xi \in \mathbb{R}.
\label{MDI:GEV}
\end{equation}
Placing independent uniform priors on $\mu$, $\log \sigma$ and $\xi$ gives the prior 
\begin{equation}
\pi_{U, GEV}(\mu,\sigma,\xi) \propto \frac{1}{\sigma}, \quad \sigma > 0, \, \mu, \xi \in \mathbb{R}. \label{eqn:unif_GEV}
\end{equation}
%which is improper for both $\sigma$ and $\xi$.
Figure \ref{fig:GEV_priors} shows the Jeffreys and MDI priors for GEV parameters as a functions of $\xi$.  The MDI prior increases without limit as $\xi \rightarrow -\infty$ and the Jeffreys prior increases without limit as $\xi \rightarrow \infty$ and as $\xi \downarrow -1/2$.

\noindent {\bf 4.2 Results}

\begin{thm}\label{th:general_GEV}
For the prior $\pi(\mu,\sigma,\xi) \propto \pi(\xi) /\sigma, \sigma>0, \mu, \xi \in \mathbb{R}$ to yield a proper posterior density function it is necessary that $n \geq 2$ and, in that event, it is sufficient that $\pi(\xi)$ is (proportional to) a proper density function.
\end{thm}

\begin{thm}\label{th:J_GEV}
There is no sample size for which the Jeffreys prior (\ref{eqn:J_GEV}) yields a proper posterior density function.
\end{thm}

Truncation of the independence Jeffreys prior to $\xi \leq \xi_{+}$ would yield a proper posterior density function if $n \geq 2$.  
In this event theorem \ref{th:general_GEV} requires only that $\int_{-1/2}^{\xi_{+}} \pi(\xi) {\rm ~d}\xi$ is finite,
where here $\pi(\xi)=\sigma \pi_{J,GEV}(\mu, \sigma, \xi)$ (see (\ref{eqn:J_GEV})).
From the proof of theorem \ref{th:J_GEV} we have 
$\pi(\xi) < 2\left[ \pi^2/6 +(1-\gamma)^2 \right]^{1/2}(1+2\xi)^{-1/2}$ for $\xi \in (-1/2,-1/2+\epsilon)$, where $\epsilon>0$.  
Therefore,
\beqnn
\int_{-1/2}^{-1/2+\epsilon} \pi(\xi) {\rm ~d}\xi
&<& 2\left[ \pi^2/6 +(1-\gamma)^2 \right]^{1/2} \int_{-1/2}^{-1/2+\epsilon} (1+2\xi)^{-1/2} {\rm ~d}\xi, \\
&=& 2^{3/2} \left[ \pi^2/6 +(1-\gamma)^2 \right]^{1/2} \epsilon^{1/2}.
\eeqnn
The integral over $(-1/2+\epsilon, \xi_{+})$ is also finite.
However, the choice of an {\it a priori} upper limit for $\xi$ may be less obvious than the choice of a lower limit.

\begin{thm}\label{th:MDI_GEV}
There is no sample size for which the MDI prior (\ref{MDI:GEV}) yields a proper posterior density function.
\end{thm}

As in the GP case, truncating the MDI prior to  $\xi \geq -1$, that is,
\begin{equation}
\pi'_{M,GEV}(\mu,\sigma,\xi) \propto \frac{1}{\sigma} \,{\rm e}^{-\gamma(1+\xi)} \quad \mu \in \mathbb{R}, \sigma > 0,  \, \xi \geq -1, 
\label{eqn:mod_GEV}
\end{equation}
is one way to yield a proper posterior distribution.

\begin{cor}\label{th:mod_MDI_GEV}
{\bf to theorem \ref{th:general_GEV}.} The truncated MDI prior (\ref{eqn:mod_GEV}) yields a proper posterior density function for $n \geq 2$.
\end{cor}

\begin{thm}\label{th:F_GEV}
A sufficient condition for the uniform prior (\ref{eqn:unif_GEV}) to yield a proper posterior density function is that $n \geq 4$.
\end{thm}

\par

%%%%%%%%%%%%%%%%%%%%%%%%%%%%%%%%%%%%%%%%%%%%%%%%%%%%%%%%%%%%%%%%%%%%%%%%%%%%%%%%%%%%%%%%%%%%%%%%%%%%%%%%%%%%%%%%%%%%%%%%%%%%

\setcounter{chapter}{5}
\setcounter{equation}{0} %-1
\noindent {\bf 5. Discussion}

We have shown that some of the reference priors used, or proposed for use, in extreme value modelling do not yield a proper posterior distribution unless we are willing to truncate the possible values of $\xi$ {\it priori}.  
An interesting aspect of our findings is that the Jeffreys prior (\ref{eqn:J_GEV}) for GEV parameters fails to yield a proper posterior, whereas the uniform prior (\ref{eqn:unif_GEV}) requires only weak conditions to ensure posterior propriety.
This is the opposite of more general experience, summarised by 
\cite[page 393]{Berger2006} and \cite[page 5]{YB1998},
that Jeffreys prior almost always yields a proper posterior whereas a uniform prior often fails to do so.
The impropriety of the posterior under the Jeffreys prior is due to the high rate at which the component $\pi(\xi)$ of this prior increases for large $\xi$.
An alternative prior based on Jeffreys' general rule (\ref{eqn:Jeff}) \citep{van2004} also has this property.

The conditions sufficient for posterior propriety under the uniform priors
(\ref{eqn:unif_GP}) and (\ref{eqn:unif_GEV}) are weak.  
Therefore, a posterior yielded by a diffuse normal priors is meaningful but such a prior could be replaced by an improper uniform prior.
Although it is reassuring to know that a posterior is proper, with a sufficiently informative sample posterior impropriety might not present a practical problem \citep[section 5.2]{Kass1996}.
This may explain why \cite[pages 435 and 447]{Beirlant2004} obtain sensible results using (untruncated) MDI priors.  However, the posterior impropriety may be evident for smaller sample sizes.

In making inferences about high quantiles of the marginal distribution of $X$, the GP model for threshold excesses is combined with a binomial($N,p_u$) model for the number of excesses, where $p_u=P(X>u)$.  Reference priors for a binomial probability have been studied extensively, see, for example, \cite{TGM2009}.
An approximately equivalent approach is the non-homogeneous Poisson process (NHPP) model \citep{Smith1989}, which is parameterized in terms of GEV parameters $\mu, \sigma$ and $\xi$ relating to the distribution of $\max\{X_, \ldots, X_b\}$.
Suppose that $m$ observations $x_1, \ldots, x_m$ exceed $u$.
Under the NHPP the posterior density for $(\mu,\sigma,\xi)$ is proportional to
\begin{equation}
\sigma^{-(m+1)} \pi(\xi)
\exp\left\{\!-n \left[ 1\!+\!\xi\left( \frac{u\!-\!\mu}{\sigma} \right)\right]_+^{-1/\xi} \right\}
%\left\{
\prod_{i=1}^m \left[ 1\!+\!\xi\left( \frac{x_i\!-\!\mu}{\sigma} \right)\right]_+^{-(1+1/\xi)},
%\right\}, 
\label{eqn:NHPP}
\end{equation}
where $n$ is the (notional) number of blocks into which the data are divided in defining $(\mu,\sigma,\xi)$.
Without loss of generality, we take $n=m$.  The exponential term in (\ref{eqn:NHPP}) is an increasing function of $u$, and $x_i>u, i=1, \ldots,m$.  Therefore, 
\[ \exp\left\{-n \left[ 1+\xi\left( \frac{u-\mu}{\sigma} \right)\right]_+^{-1/\xi} \right\} < \exp\left\{-\sum_{i=1}^m \left[ 1+\xi\left( \frac{x_i-\mu}{\sigma} \right)\right]_+^{-1/\xi} \right\} \]
and (\ref{eqn:NHPP}) is less than
\begin{equation}
\sigma^{-(m+1)} \pi(\xi)
\exp\left\{\!-\sum_{i=1}^m \!\left[ 1\!+\!\xi\left(\!\frac{x_i\!-\!\mu}{\sigma}\!\right)\right]_+^{-1/\xi} \right\}
%\left\{
\prod_{i=1}^m \left[ 1\!+\!\xi\left(\!\frac{x_i\!-\!\mu}{\sigma}\!\right)\right]_+^{-(1+1/\xi)}\!\!.
%\right\}.
\label{eqn:NHPPnewer}
\end{equation}
Equation (\ref{eqn:NHPPnewer}) is of the same form as (\ref{eqn:GEVpost}), with $n=m$ and $y_i=x_i, i=1, \ldots, n$.  Therefore, theorems \ref{th:general_GEV} and \ref{th:F_GEV} apply to the NHPP model, that is, for posterior propriety it is sufficient that either (a) $n \geq 2$ and $\pi(\mu,\sigma,\xi) \propto \pi(\xi)/\sigma$, for $\sigma > 0, \mu, \xi \in \mathbb{R}$, where $\int_\xi \pi(\xi) {\rm~d}\xi$ is finite, or (b)
 $n \geq 4$ and $\pi(\mu,\sigma,\xi) \propto 1/\sigma$, for $\sigma > 0, \mu, \xi \in \mathbb{R}$.
%SmithGoodman

One possible extension of our work is to regression modelling using extreme value response distributions.  For example, \cite{RD2014} use GEV regression modelling to analyze reliability data.  They prove posterior propriety under conditions on the prior for $(\sigma,\xi)$ that are stronger than those in our theorems \ref{th:general_GEV} and \ref{th:F_GEV}.
Future work will investigate our conjecture that the conditions in \cite{RD2014} can be weakened.
Another extension is to explore other formal rules for constructing priors, such as reference priors \citep{BBS2009} and probability matching priors \citep{DMGS2000}.
\cite{Ho2010} considers the latter for the GP distribution.
\par

%%%%%%%%%%%%%%%%%%%%%%%%%%%%%%%%%%%%%%%%%%%%%%%%%%%%%%%%%%%%%%%%%%%%%%%%%%%%%%%%%%%%%%%%%%%%%%%%%%%%%%%%%%%%%%%%%%%%%%%%%%%%

\noindent {\large\bf Acknowledgements}

One of us (NA) was funded by an Engineering and Physical Sciences Research
Council studentship while carrying out this work.   
%We are grateful to two referees and an associate editor for their constructive comments on an earlier draft.
\par

%%%%%%%%%%%%%%%%%%%%%%%%%%%%%%%%%%%%%%%%%%%%%%%%%%%%%%%%%%%%%%%%%%%%%%%%%%%%%%%%%%%%%%%%%%%%%%%%%%%%%%%%%%%%%%%%%%%%%%%%%%%%

%%%%%%%%%%%%%%%%%%%%%%%%%%%%%%%%%%%%%%%%%%%%%%%%%%%%%%%%%%%%%%%%%%%%%%%%%%%%%%%%%%%%%%%%%%%%%%%%%%%%%%%%%%%%%%%%%%%%%%%%%%%%

%\pagebreak

\setcounter{chapter}{6}
\setcounter{equation}{0} %-1
\noindent {\bf 6. Appendix}

\noindent {\bf 6.1 Moments of a GP distribution}

We give some moments of the GP distribution for later use.
Suppose that $Z \sim GP(\sigma,\xi)$, where $\xi<1/r$.  Then
\citep{GF2009}
\beqn
{\rm E}(Z^r)&=&\frac{r! \, \sigma^r}{\displaystyle\prod_{i=1}^r (1-i \xi)}, \qquad r= 1, 2, \ldots. 
\label{eqn:GPmomr}
\eeqn
Now suppose that $\xi<0$.
Then, for a constant $a > \xi$, and using the substitution $x=-\xi v/\sigma$, we have
\beqn
{\rm E}(Z^{-a/\xi})
&=&
\int_0^{-\sigma/\xi} v^{-a/\xi} \frac{1}{\sigma}
\left( 1+\frac{\xi v}{\sigma} \right)^{-(1+1/\xi)} {\rm ~d}v, \nonumber \\
&=&
(-\xi)^{a/\xi-1} \sigma^{-a/\xi} \int_0^1 x^{-a/\xi} (1-x)^{-(1+1/\xi)} {\rm ~d}x, \nonumber \\
&=&
(-\xi)^{a/\xi-1} \sigma^{-a/\xi} \frac{\Gamma(1-a/\xi) \Gamma(-1/\xi)}{\Gamma(1-(a+1)/\xi)},
\label{eqn:GPmom}
\eeqn
where we have used integral number 1 in section 3.251 on page 324 of \cite{GR1965}, namely
\[ \int_0^1 x^{\mu-1} (1-x^\lambda)^{\nu-1} {\rm ~d}x 
= \frac{1}{\lambda} \mbox{Beta}\left(\frac{\mu}{\lambda},\nu\right) = \frac{\Gamma(\mu/\lambda) \Gamma(\nu)}{\Gamma(\mu/\lambda+\nu)} \qquad \lambda>0, \nu>0, \mu>0, \]
with $\lambda=1, \mu=1-a/\xi$ and $v=-1/\xi$.

In the following proofs we use the generic notation $\pi(\xi)$ for the component of the prior relating to $\xi$: the form of $\pi(\xi)$ varies depending on the prior being considered.

%%%%%%%%%%%%%%%%%%%%%%%%%%%%%%%%%%%%%%%%%%%%%%%%%%%%%%%%%%%%%%%%%%%%%%%%%%%%%%%%%%%%%%%%%%%%%%%%%%%%%%%%%%%%%%%%%%%%%%%%%%%%

\noindent {\bf 6.2 Proof of theorem \ref{th:general_GP}  and its corollary}

This trivial extension of the proof of theorem 1 in \cite{EugeniaCastellanos2007}.
Suppose $m=1$, with an observation $z$.  The normalizing constant $C$ of the posterior distribution is given by
\beqnn
C_1\!\!\!&=&\!\!\! \int_{-\infty}^{0} \!\!\!\pi(\xi) \int_{-\xi z}^\infty 
\!\!\!\sigma^{-2}\!\left( 1+\xi z / \sigma \right)^{-(1+1/\xi)} \!{\rm ~d}\sigma {\rm d}\xi
+ \!\int_{0}^{\infty} \!\!\!\pi(\xi) \int_{0}^\infty 
\!\!\!\sigma^{-2}\!\left( 1+\xi z / \sigma \right)^{-(1+1/\xi)} \!{\rm ~d}\sigma {\rm d}\xi, \\
\!\!\!&=&\!\!\! \frac{1}{z} \int_{-\infty}^\infty \pi(\xi) {\rm ~d}\xi. \\
\eeqnn
If the latter integral is finite, that is, $\pi(\xi)$ is proportional to a proper density function, then the posterior distribution is proper for $m=1$ and therefore, by successive iterations of Bayes' theorem, it is proper for $m \geq 1$.  

The corollary follows directly.
\qed

%%%%%%%%%%%%%%%%%%%%%%%%%%%%%%%%%%%%%%%%%%%%%%%%%%%%%%%%%%%%%%%%%%%%%%%%%%%%%%%%%%%%%%%%%%%%%%%%%%%%%%%%%%%%%%%%%%%%%%%%%%%%

\noindent {\bf 6.3 Proof of theorem \ref{th:MDI_GP}}

Let $A(\xi)={\rm e}^{-\xi}$ and $B(\sigma, \xi)=\sigma^{-(m+1)} \prod_{i=1}^m
\left( 1+\xi z_i / \sigma \right)^{-(1+1/\xi)}$.  
Then, from (\ref{eqn:const_GP}) we have
\beqnn
C_m &=& \int_{-\infty}^\infty A(\xi) \int_{\max(0,-\xi z_m)}^\infty \!\!\!B(\sigma, \xi) {\rm ~d}\sigma {\rm d}\xi, \\
&=& \int_{-\infty}^{-1} \!\!\!\!A(\xi) \int_{-\xi z_m}^\infty \!\!\!\!\!B(\sigma, \xi) {\rm ~d}\sigma {\rm d}\xi 
+ \int_{-1}^{0} \!\!\!\!A(\xi) \int_{-\xi z_m}^\infty \!\!\!\!\!B(\sigma, \xi) {\rm ~d}\sigma {\rm d}\xi
+ \int_{0}^{\infty} \!\!\!\!A(\xi) \int_{0}^\infty \!\!\!\!\!B(\sigma, \xi) {\rm ~d}\sigma {\rm d}\xi.
\eeqnn
The latter two integrals converge for $m \geq 1$.
However, the first integral diverges for all samples sizes.
For $\xi < -1$, $(1+\xi z / \sigma)^{-(1+1/\xi)} > 1$ when $z$ is in the support  $(0,-\sigma/\xi)$ of the GP($\sigma,\xi)$ density.  Therefore $B(\sigma,\xi) > \sigma^{-(m+1)}$.
Thus, the first integral above satisfies
\beqnn
\int_{-\infty}^{-1} A(\xi) \int_{-\xi z_m}^\infty B(\sigma, \xi) {\rm ~d}\sigma {\rm ~d}\xi &>&
\int_{-\infty}^{-1} A(\xi) \int_{-\xi z_m}^\infty \sigma^{-(m+1)} {\rm ~d}\sigma {\rm ~d}\xi, \\
&=& \int_{-\infty}^{-1} A(\xi) \left[ -\frac1m \sigma^{-m} \right]_{-\xi z_m}^\infty {\rm ~d}\xi, \\
&=& \int_{-\infty}^{-1} A(\xi) \frac1m  \left[ - \xi z_m \right]^{-m} {\rm ~d}\xi, \\
&=& \frac{1}{m z_m^{m}} \int_1^\infty v^{-m} {\rm e}^v {\rm ~d}v,
\eeqnn
where $v=-\xi$.  This integral is divergent for all $m \geq 1$, so there is no sample size for which the posterior is proper.
\qed

%%%%%%%%%%%%%%%%%%%%%%%%%%%%%%%%%%%%%%%%%%%%%%%%%%%%%%%%%%%%%%%%%%%%%%%%%%%%%%%%%%%%%%%%%%%%%%%%%%%%%%%%%%%%%%%%%%%%%%%%%%%%

\noindent {\bf 6.4 Proof of theorem \ref{th:F_GP}}

We need to show that $C_3$ is finite.  
We split the range of integration over $\xi$ so that $C_3=I_1+I_2+I_3$, where
\begin{equation}
I_1=\int_{-\infty}^{-1} \int_{-\xi z_3}^{\infty} \!B(\sigma,\xi) {\rm ~d}\sigma \!{\rm ~d}\xi, \quad
I_2=\int_{-1}^{0} \int_{-\xi z_3}^{\infty} \!B(\sigma,\xi) {\rm ~d}\sigma \!{\rm ~d}\xi, \quad
I_3=\int_{0}^{\infty} \!\int_{0}^{\infty} \!B(\sigma,\xi) {\rm ~d}\sigma \!{\rm ~d}\xi \nonumber
\end{equation}
and $B(\sigma, \xi)=\sigma^{-4} \prod_{i=1}^3 \left( 1+\xi z_i / \sigma \right)^{-(1+1/\xi)}$.
For convenience we let $\rho=\xi/\sigma$.

\noindent {\bf Proof that $I_1$ is finite.}
\newcommand{\y}[1]{(1+\rho z_{#1})}
\newcommand{\yx}[1]{\left(1+\frac{\xi z_{#1}}{\sigma} \right)}
\newcommand{\yy}[2]{(z_{#1}-z_{#2})}
We have $\xi < -1$ and so $-(1+1/\xi)<0$, $\rho<0$ and $0<1+\rho z_i<1$ for $i=1,2,3$.  Noting that $-\rho z_3 < 1$ gives
\beqn
(1+\rho z_1)(1+\rho z_2)(1+\rho z_3) &>& 
(-\rho z_3+\rho z_1)(-\rho z_3+\rho z_2)(1+\rho z_3), \nonumber \\
&=& (-\rho)^2(z_3-z_1)(z_3-z_2)(1+\rho z_3), \nonumber \\
&=& (-\xi)^2\sigma^{-2} (z_3-z_1)(z_3-z_2)(1+\rho z_3). \label{eqn:mine}
\eeqn
Therefore,
\[ \prod_{i=1}^3 \left( 1+\frac{\xi z_i}{\sigma} \right)^{\!\!-(1+1/\xi)} <
(-\xi)^{-2(1+1/\xi)} \sigma^{2(1+1/\xi)} \left[\yy{3}{2}\yy{3}{1} \yx{3}\right]^{\!-(1+1/\xi)}.  \] 
Thus,
\beqnn
I_1 &\leq& 
\int_{-\infty}^{-1} (-\xi)^{-2(1+1/\xi)} \left[\yy{3}{2}\yy{3}{1} \right]^{\!-(1+1/\xi)}
I_{1\sigma} \!{\rm ~d}\xi, 
\eeqnn
where
\beqnn
I_{1\sigma}&=&
\int_{-\xi z_3}^{\infty}
\sigma^{-4} \sigma^{2(1+1/\xi)}  \yx{3}^{\!-(1+1/\xi)}
{\rm ~d}\sigma, \\
&=& z_3^{-1} \int_{0}^{-1/\xi z_3} v^{-2/\xi} 
\frac{1}{z_3^{-1}} \left(1+\frac{\xi v}{z_3^{-1}}\right)^{\!\!-(1+1/\xi)} {\rm ~d}v, \\
&=& (-\xi)^{2/\xi-1} z_3^{-(1-2/\xi)} \, \frac{\Gamma(1-2/\xi) \Gamma(-1/\xi)}{\Gamma(1-3/\xi)}, 
\eeqnn
where $v=1/\sigma$ and the last line follows from (\ref{eqn:GPmom}) with $a=2$ and $\sigma=z_3^{-1}$.
%on noting that the integrand is ${\rm E}\left(V^{-2/\xi}\right)$, where $V \sim \mbox{GP}(z_3^{-1},\xi)$.
Therefore,
\beqn
I_1 &\leq& 
\int_{-\infty}^{-1} (-\xi)^{-3} \left[\yy{3}{2}\yy{3}{1} \right]^{\!-(1+1/\xi)}
z_3^{-(1-2/\xi)} \frac{\Gamma(1-2/\xi) \Gamma(-1/\xi)}{\Gamma(1-3/\xi)}  \!{\rm ~d}\xi, \nonumber \\
&=& 
[z_3 \yy{3}{2} \yy{3}{1}]^{-1} \int_{-\infty}^{-1} (-\xi)^{-3} 
\left( 1-\frac{z_2}{z_3}\right)^{\!\!-1/\xi} \left( 1-\frac{z_1}{z_3}\right)^{\!\!-1/\xi}
\frac{\Gamma(1-2/\xi) \Gamma(-1/\xi)}{\Gamma(1-3/\xi)}  \!{\rm ~d}\xi, \nonumber  \\
&=& 
[z_3 \yy{3}{2} \yy{3}{1}]^{-1} \int_{0}^{1} x 
\left( 1-\frac{z_2}{z_3}\right)^{\!\!x} \left( 1-\frac{z_1}{z_3}\right)^{\!\!x}
\frac{\Gamma(1+2x) \Gamma(x)}{\Gamma(1+3x)}  \!{\rm ~d}x, \nonumber \\
&=& 
[z_3 \yy{3}{2} \yy{3}{1}]^{-1} \int_{0}^{1} 
\left( 1-\frac{z_2}{z_3}\right)^{\!\!x} \left( 1-\frac{z_1}{z_3}\right)^{\!\!x}
\frac{\Gamma(1+2x) \Gamma(1+x)}{\Gamma(1+3x)}  \!{\rm ~d}x, 
\label{eqn:finite}
\eeqn
where $x=-1/\xi$ and we have used the relation $\Gamma(1+x)=x\,\Gamma(x)$.
The integrand in (\ref{eqn:finite}) is finite over the range of integration so this integral is finite and therefore $I_1$ is finite. 

\noindent {\bf Proof that $I_2$ is finite.}
We have $-1 < \xi < 0$, so $-(1+1/\xi)>0$ and 
%\linebreak[4] 
$(1+\xi z /\sigma)^{-(1+1/\xi)}<1$ and decreases in $z$ over $(0,-\sigma/\xi)$.
Therefore,
\beqnn
I_2&=& \int_{-1}^{0} \int_{-\xi z_3}^{\infty} 
\sigma^{-4} \prod_{i=1}^3 \left( 1+\frac{\xi z_i}{\sigma} \right)^{-(1+1/\xi)}
{\rm ~d}\sigma {\rm ~d}\xi, \\
&\leq& 
\int_{-1}^{0} \int_{-\xi z_3}^{\infty} 
\sigma^{-4} \left( 1+\frac{\xi z_3}{\sigma} \right)^{-(1+1/\xi)}
{\rm ~d}\sigma {\rm ~d}\xi, \\
&=&
\int_{-1}^{0} z_3^{-1} \int_0^{-1/\xi z_3} 
v^{2} \frac{1}{z_3^{-1}} \left( 1+\frac{\xi v}{z_3^{-1}} \right)^{-(1+1/\xi)}
{\rm ~d}v {\rm ~d}\xi, \\
&=& 
 z_3^{-1}  \int_{-1}^{0} \frac{2 z_3^{-2}}{(1-\xi)(1-2\xi)}  {\rm ~d}\xi, \\
&=& 
 2 z_3^{-3}  \int_{-1}^{0} \left\{ \left(\frac12-\xi\right)^{-1} - \left(1-\xi\right)^{-1}   \right\}  {\rm ~d}\xi, \\
&=& 2 z_3^{-3} \ln(3/2),
\eeqnn
where the integral over $v$ follows from (\ref{eqn:GPmomr}) with $r=2$ and $\sigma=z_3^{-1}$.

\noindent {\bf Proof that $I_3$ is finite.}
We have $\xi >0$ so $-(1+1/\xi)<0$.  
Let $g_n = (\prod_{i=1}^n z_i)^{1/n}$.
\citet[page 130]{Mitrinovic1964}:
\begin{equation}
\prod_{k=1}^n (1+a_k) \geq (1+b)^n, \qquad a_k >0; \quad \prod_{k=1}^n a_k = b^n,  \label{eqn:mit}
\end{equation}
with $a_k=\xi z_k/\sigma$ and $b=\xi g_3/\sigma$ gives
\[ \prod_{i=1}^3 \left(1+\frac{\xi z_i}{\sigma}\right)^{-(1+1/\xi)} 
\leq \left(1+\frac{\xi g_3}{\sigma}\right)^{-3(1+1/\xi)}, \]
and therefore
\beqnn
I_3&=&\int_{0}^{\infty} \int_{0}^{\infty} \sigma^{-4} 
\prod_{i=1}^{3}
\left( 1+\frac{\xi z_i}{\sigma} \right)^{-(1+1/\xi)}
{\rm ~d}\sigma \!{\rm ~d}\xi, \\
&\leq& 
\int_{0}^{\infty} \int_{0}^{\infty} \sigma^{-4} 
\left( 1+\frac{\xi g_3}{\sigma} \right)^{-3(1+1/\xi)}
{\rm ~d}\sigma \!{\rm ~d}\xi, \\
&=&
\int_{0}^{\infty} \beta \int_{0}^{\infty} v^2
\frac{1}{\beta} \left( 1+\frac{\alpha v}{\beta} \right)^{-(1+1/\alpha)}
{\rm ~d}v \!{\rm ~d}\xi, 
\eeqnn
where $v=1/\sigma$, $\alpha=1/(2+3/\xi)$ and $\beta=\alpha/\xi g_3=1/(3+2\xi)g_3$.
For $\xi>0$, $\alpha < 1/2$ so using (\ref{eqn:GPmomr}) with $r=2, \sigma=\beta$ and $\xi=\alpha$ gives
\beqnn
I_3 &\leq& 
\int_{0}^{\infty} \beta  \frac{2 \beta^2}{(1-\alpha)(1-2\alpha)} {\rm ~d}\xi, \\
&=&
\frac{2}{3} g_3^{-3} \int_{0}^{\infty} \frac{1}{(\xi+3)(2\xi+3)} {\rm ~d}\xi, \\
&=&
\frac{2}{9} g_3^{-3} \int_{0}^{\infty} \left( \frac{1}{\xi+3/2} - \frac{1}{\xi+3} \right) {\rm ~d}\xi, \\
&=&
\frac{2}{9} g_3^{-3} \ln 2.
\eeqnn
The normalizing constant $C_3$ is finite, so $\pi_{U,GP}(\sigma,\xi)$ yields a proper posterior density for $m=3$ and therefore does so for $m \geq 3$.
\qed

%%%%%%%%%%%%%%%%%%%%%%%%%%%%%%%%%%%%%%%%%%%%%%%%%%%%%%%%%%%%%%%%%%%%%%%%%%%%%%%%%%%%%%%%%%%%%%%%%%%%%%%%%%%%%%%%%%%%%%%%%%%%

\noindent {\bf 6.5 Proof of theorem \ref{th:general_GEV} and its corollary}

Throughout the following proofs we define $\delta_i=y_i-y_1, i=2, \ldots, n$.

We make the parameter transformation $\phi=\mu-\sigma/\xi$.  Then the posterior density for $(\phi,\sigma,\xi)$ is given by
\[ \pi(\phi,\sigma,\xi) = K_n^{-1} \pi(\xi)
%\e^{-\gamma(1+\xi)}  
|\xi|^{-n(1+1/\xi)} G_n(\phi,\sigma), \]
where
\[ G_n(\phi,\sigma) = \sigma^{n/\xi-1} \left\{\prod_{i=1}^n |y_i-\phi|^{-(1+1/\xi)}\right\} \exp\left\{-|\xi|^{-1/\xi} \, \sigma^{1/\xi} \sum_{i=1}^n |y_i-\phi|^{-1/\xi} \right\} \]
and, if $\xi >0$ then $\phi < y_1$ and if $\xi < 0$ then $\phi > y_n$.

We let $b=|\xi|^{-1/\xi} \sum_{i=1}^n |y_i-\phi|^{-1/\xi}$ and $v=\sigma^{1/\xi}$.
The normalizing constant $K_n$ is given by 
\beqn
K_n &=& 
\int_{-\infty}^\infty \int \int_0^\infty \pi(\xi) |\xi|^{-n(1+1/\xi)} G_n(\phi,\sigma)
{\rm ~d}\sigma {\rm ~d}\phi {\rm ~d}\xi, \nonumber \\
&=& 
\int_{-\infty}^\infty \pi(\xi) |\xi|^{-n(1+1/\xi)}  \int  \left\{\prod_{i=1}^n |y_i-\phi|^{-(1+1/\xi)}\right\}  \int_0^\infty 
\sigma^{n/\xi-1} 
\exp\left\{-b \sigma^{1/\xi} \right\}
{\rm ~d}\sigma {\rm ~d}\phi {\rm ~d}\xi, \nonumber \\
&=& 
\int_{-\infty}^\infty \pi(\xi) |\xi|^{-n(1+1/\xi)}  \int  \left\{\prod_{i=1}^n |y_i-\phi|^{-(1+1/\xi)}\right\}  \int_0^\infty 
v^{n-1} \exp\{-b v\} 
\,|\xi| {\rm ~d}v {\rm ~d}\phi {\rm ~d}\xi, \nonumber \\
&=& 
\int_{-\infty}^\infty \pi(\xi) |\xi|^{-n(1+1/\xi)}  \int  \left\{\prod_{i=1}^n |y_i-\phi|^{-(1+1/\xi)}\right\} 
\Gamma(n) b^{-n}
\,|\xi| {\rm ~d}\phi {\rm ~d}\xi, \nonumber \\
&=&
\int_{-\infty}^\infty \pi(\xi) |\xi|^{-n(1+1/\xi)}  \int  \left\{\prod_{i=1}^n |y_i-\phi|^{-(1+1/\xi)}\right\} 
(n-1)! |\xi|^{n/\xi+1} \left\{ \sum_{i=1}^n |y_i-\phi|^{-1/\xi} \right\}^{\!\!-n}
{\rm ~d}\phi {\rm ~d}\xi, \nonumber \\
&=&
(n-1)! \int_{-\infty}^\infty \pi(\xi) |\xi|^{1-n}  \int  \left\{\prod_{i=1}^n |y_i-\phi|^{-(1+1/\xi)}\right\} 
\left\{ \sum_{i=1}^n |y_i-\phi|^{-1/\xi} \right\}^{\!\!-n}
{\rm ~d}\phi {\rm ~d}\xi, \label{eqn:GEVa}
\eeqn

For $n=1$ the integral $\int_{\phi: \xi(y_1-\phi)>0} |y_1-\phi|^{-1} {\rm ~d}\phi$ is divergent so if $n=1$ the posterior is not proper for any prior in this class.

Now we take $n=2$ and for clarity consider the cases $\xi>0$ and $\xi<0$ separately, with respective contributions $K_2^+$ and $K_2^-$ to $K_2$.
For $\xi>0$, using the substitution $u=(y_1-\phi)^{-1}$ in (\ref{eqn:GEVa}) gives
\beqnn
K_2^+
&=&
\int_{0}^\infty \pi(\xi) \,\xi^{-1}  
\int_{-\infty}^{y_1} 
\frac{(y_1-\phi)^{-(1+1/\xi)}(y_2-\phi)^{-(1+1/\xi)}}
{\left\{(y_1-\phi)^{-1/\xi}+(y_2-\phi)^{-1/\xi}\right\}^2} {\rm ~d}\phi {\rm ~d}\xi,  \\
&=& \int_{0}^\infty \pi(\xi) \, \xi^{-1}  
\int_0^{\infty} 
\frac{(1+\delta_2 u)^{-(1+1/\xi)}}
 {\left\{1+(1+\delta_2 u)^{-1/\xi}\right\}^2}  {\rm ~d}u {\rm ~d}\xi,   \\
&=& \frac12 \delta_2^{-1} \int_{0}^\infty \pi(\xi) {\rm ~d}\xi,   
\eeqnn
the final step following because the $u$-integrand is a multiple ($\xi \delta_2^{-1}$) of a shifted log-logistic density function with location, scale and shape parameters of $0, \xi \delta_2^{-1}$ and $\xi$ respectively, and the location of this distribution equals the median.
For $\xi <0$ an analogous calculation using the substitution $v=(y_n-\phi)^{-1}$ in (\ref{eqn:GEVa}) gives
\beqnn
K_2^- &=& \frac12 \delta_2^{-1} \int_{-\infty}^0 \pi(\xi) {\rm ~d}\xi.
\eeqnn
Therefore,
\beqnn
K_2 &=& K_2^++K_2^- = \frac12 \delta_2^{-1} \int_{-\infty}^{\infty} \pi(\xi) {\rm ~d}\xi.
\eeqnn
Thus, $K_2$ is finite if $\int_{-\infty}^{\infty} \pi(\xi) {\rm ~d}\xi$ is finite, and the result follows.

The corollary follows directly.
\qed

%%%%%%%%%%%%%%%%%%%%%%%%%%%%%%%%%%%%%%%%%%%%%%%%%%%%%%%%%%%%%%%%%%%%%%%%%%%%%%%%%%%%%%%%%%%%%%%%%%%%%%%%%%%%%%%%%%%%%%%%%%%%

\noindent {\bf 6.6 Proof of theorem \ref{th:J_GEV}}

The crucial aspects are the rates at which $\pi(\xi)$ $\rightarrow \infty$ as $\xi \downarrow -1/2$ and as $\xi \rightarrow \infty$.  

The component $\pi(\xi)$ of (\ref{eqn:J_GEV}) involving $\xi$ can be expressed as \begin{equation}
\pi^2_\xi(\xi)=\frac{1}{\xi^4}(T_1 + T_2), \label{my_Jeff}
\end{equation}
where
\beqn
T_1&=& \left[ \frac{\pi^2}{6} +(1-\gamma)^2 \right] (1+\xi)^2 \, \Gamma(1+2\xi), \label{JGEVa} \\
T_2&=& \frac{\pi^2}{6}+\left[ 2(1-\gamma)(\gamma+\psi(1+\xi))-\frac{\pi^2}{3} \right]\Gamma(2+\xi), \nonumber \\
&&  -\left[1+\psi(1+\xi)\right]^2 \left[\Gamma(2+\xi)\right]^2. \label{JGEVb}
\eeqn
Firstly, we derive a lower bound for $\pi(\xi)$ that holds for $\xi > 3$.
Using the duplication formula \citep[page 256; 6.1.18]{AS1972} 
\[ \Gamma(2z) = (2\pi)^{-1/2} \, 2^{\,2z-1/2} \, \Gamma(z) \, \Gamma(z+1/2), \]
with $z=1/2+\xi$ in (\ref{JGEVa}) we have
\beqnn
T_1&=&\left[ \frac{\pi^2}{6} +(1-\gamma)^2 \right] (1+\xi)^2 \, \pi^{-1/2} 2^{2\xi} \, \Gamma(1/2+\xi) \, \Gamma(1+\xi).
\eeqnn
We note that
\[ \Gamma(1/2+\xi) =\frac{\Gamma(3/2+\xi)}{1/2+\xi} > \frac{\Gamma(1+\xi)}{1/2+\xi} = \frac{2\Gamma(1+\xi)}{1+2\xi} > \frac{\Gamma(1+\xi)}{1+\xi}, \]
where for the first inequality to hold it is sufficient that $\xi > 1/2$; and that, for $\xi > 3$, $2^{2\xi} > (1+\xi)^3$.
Therefore,
\begin{equation}
T_1 > \left[ \frac{\pi^2}{6} +(1-\gamma)^2 \right] \pi^{-1/2} \, (1+\xi)^4 \, [\Gamma(1+\xi)]^2. \label{T1}
\end{equation}

Completing the square in (\ref{JGEVb}) gives
\beqnn
T_2 &=&-\left\{ [1+\psi(1+\xi)]\,\Gamma(2+\xi) + f(\xi) \right\}^2+[f(\xi)]^2+\pi^2/6,
\eeqnn
where 
\[ f(\xi) = \frac{\pi^2/6-(1-\gamma)(\gamma+\psi(1+\xi))}{1+\psi(1+\xi)} 
= \frac{\pi^2/6+(1-\gamma)^2}{1+\psi(1+\xi)} - (1-\gamma)\]
and $[f(\xi)]^2+\pi^2/6>0$.

For $\xi>0$, $\psi(1+\xi)$ increases with $\xi$ and so $f(\xi)$ decreases with $\xi$.
Therefore, for $\xi > 3$, $f(\xi)<f(3)\approx 0.39$ and 
\beqnn
T_2 &>& -\left\{ [1+\psi(1+\xi)]\,\Gamma(2+\xi) + f(3) \right\}^2.
\eeqnn
For $\xi>0$, we have $\psi(1+\xi) < \ln(1+\xi)-(1+\xi)^{-1}/2$ \citep[theorem C]{QV2004} 
and $\ln(1+\xi) \leq \xi$ \citep[page 68; 4.1.33]{AS1972}.
Therefore, noting that $\Gamma(2+\xi)=(1+\xi)\,\Gamma(1+\xi)$ we have 
\beqnn
T_2 &>& -\left\{ (1+\xi)^2\,\Gamma(1+\xi) -\frac12\Gamma(1+\xi) + f(3) \right\}^2.
\eeqnn
For $\xi>3$, $f(3)-\Gamma(1+\xi)/2<0$ so
\begin{equation}
T_2 > -(1+\xi)^4 \, [\Gamma(1+\xi)]^2. \label{T2}
\end{equation}
Substituting (\ref{T1}) and (\ref{T2}) in (\ref{my_Jeff}) gives, for $\xi>3$,
\beqnn
\pi^2_\xi(\xi) &>&  \frac{(1+\xi)^4}{\xi^4}
\left\{
\left[ \frac{\pi^2}{6} +(1-\gamma)^2 \right] \pi^{-1/2} - 1
\right\}
 [\Gamma(1+\xi)]^2, \nonumber \\
&>& c [\Gamma(1+\xi)]^2, \\
&>& c(1+\xi)^{2(\lambda\xi-\gamma)},
\eeqnn
where $c=(4/3)^4 \{[ \pi^2/6 +(1-\gamma)^2 ] \pi^{-1/2} - 1\} \approx 0.0913$
and the final step uses the inequality 
$\Gamma(x) > x^{\lambda(x-1)-\gamma}$, for $x>0$ \citep{Alzer1999},
where $\lambda=(\pi^2/6-\gamma)/2 \approx 0.534$.
Thus, a lower bound for the $\xi$ component of the Jeffreys prior (\ref{eqn:J_GEV}) is given by
\beqn
\pi(\xi) &>& c^{1/2} (1+\xi)^{\lambda \xi - \gamma}, \qquad \mbox{for~} \xi>3. 
\label{eqn:finally}
\eeqn
[In fact, numerical work shows that this lower bound holds for $\xi > -1/2$.]

Let $K_n^+$ denote the contribution to $K_n$ for $\xi>3$.
Using the substitution $u=(y_1-\phi)^{-1}$ in (\ref{eqn:GEVa}) gives
\beqn
K_n^+
&\!\!\!=\!\!\!&
(n\!-\!1)! \int_{3}^\infty \!\!\!\pi(\xi) \,\xi^{1-n}  
\!\int_0^{\infty} \!\!\!u^{n-2} 
\frac{\displaystyle\prod_{i=1}^n (1+\delta_i u)^{-(1+1/\xi)}}
{\left\{ 1+\displaystyle\sum_{i=2}^n (1+\delta_i u)^{-1/\xi} \right\}^n} 
{\rm ~d}u\!
{\rm ~d}\xi. \label{eqn:GEVb}
\eeqn
For $\xi>0$ we have $1+\displaystyle\sum_{i=2}^n (1+\delta_i u)^{-1/\xi}  \leq n$ and $\prod_{i=1}^n (1+\delta_i u)^{-(1+1/\xi)} \geq$ 
%\linebreak[4] 
$(1+\delta_n u)^{-(n-1)(1+1/\xi)}$.  Applying these inequalities to (\ref{eqn:GEVb}) gives
\beqn
K_n^+
&\!\!\!\geq\!\!\!&
n^{-n} (n-1)! \int_{3}^\infty \pi(\xi) \,\xi^{1-n}  
\int_0^{\infty} u^{n-2} (1+\delta_n u)^{-(n-1)(1+1/\xi)}
{\rm ~d}u
{\rm ~d}\xi, \nonumber \\
&\!\!\!=\!\!\!& 
n^{-n} (n\!-\!1)! \int_{3}^\infty \!\!\!\pi(\xi) \,\xi^{1-n}  
\beta \int_0^{\infty} \!\!\!u^{n-2} \frac{1}{\beta} \left(1+\frac{\alpha u}{\beta}\right)^{-(1+1/\alpha)}
{\rm ~d}u
{\rm ~d}\xi, \label{eqn:Knplus}
\eeqn
where $\beta=\alpha/\delta_n$ and $\alpha=[n-2+(n-1)/\xi]^{-1}$ and $0<\alpha<(n-2)^{-1}$.
The $u$-integrand is the density function of a GP($\beta,\alpha$) distribution 
and so, using (\ref{eqn:GPmomr}) with $r=n-2$, the integral over $u$ is given by 
\beqn
(n-2)!\,\beta^{n-2}\prod_{i=1}^{n-2} \frac{1}{1-i \alpha}
= (n-2)! \, \xi^{n-2}\delta_n^{2-n}
\prod_{i=1}^{n-2} \frac{1}{(n-2-i)\xi+n-1}. \label{eqn:GPmoment}
\eeqn
Substituting (\ref{eqn:GPmoment}) into (\ref{eqn:Knplus}) gives
\beqnn
K_n^+ &\geq& 
n^{-n} (n-1)! (n-2)! \, \delta_n^{1-n} \int_{3}^\infty 
\frac{1}{(n-2)\xi+n-1}
\prod_{i=1}^{n-2} \frac{1}{(n-2-i)\xi+n-1}
\, \pi(\xi) {\rm ~d}\xi, \\
&=& 
n^{-n} (n-1)! (n-2)! \, \delta_n^{1-n} \int_{3}^\infty 
\prod_{i=0}^{n-2} \frac{1}{(n-2-i)\xi+n-1}
\, \pi(\xi) {\rm ~d}\xi, \\
&=& 
n^{-n} (n-1)! (n-2)! \, \delta_n^{1-n} (n-1)^{1-n}
\int_{3}^\infty 
\prod_{i=0}^{n-2} \frac{1}{1+\frac{i}{n-1}\xi}
\, \pi(\xi) {\rm ~d}\xi, \\
&>& 
C(n) \int_{3}^\infty 
\frac{1}{(1+\xi)^{n-2}} \, \pi(\xi) {\rm ~d}\xi, \\
\eeqnn
where $C(n)=n^{-n} (n-1)! (n-2)! \, \delta_n^{1-n} (n-1)^{1-n}$.
Applying (\ref{eqn:finally}) gives
\beqnn
K_n^+ &>& C(n) \, c^{1/2} \int_{3}^\infty 
(1+\xi)^{2-n + \lambda \xi - \gamma} {\rm ~d}\xi. 
\eeqnn
For any sample size $n$ the integrand $\rightarrow \infty$ as $\xi \rightarrow \infty$.  Therefore, the integral diverges and the result follows.
\qed

Now we derive an upper bound for $\pi_{\xi}(\xi)$ that applies for $\xi$ close to $-1/2$.
We note that for $-1/2 < \xi < 0$ we have $\Gamma(1+2\xi)=\Gamma(2+2\xi)/(1+2\xi) < (1+2\xi)^{-1}$.  From (\ref{my_Jeff}) we have
\beqnn
\pi^2_\xi(\xi)
&=& \left[ \frac{\pi^2}{6} +(1-\gamma)^2 \right] \left(\frac{1+\xi}{\xi^2}\right)^2 \Gamma(1+2\xi) + \frac{T_2}{\xi^4},
\eeqnn
where $T_2 \rightarrow -3.039$ as $\xi \downarrow -1/2$.
Noting that $(1+\xi)^2/\xi^4 \rightarrow 4$ as $\xi \downarrow -1/2$ shows that $\pi(\xi) < 2\left[ \pi^2/6 +(1-\gamma)^2 \right]^{1/2}(1+2\xi)^{-1/2}$ for $\xi \in (-1/2,-1/2+\epsilon)$, for some $\epsilon>0$.  In fact numerical work shows that $\epsilon \approx 1.29$.

%%%%%%%%%%%%%%%%%%%%%%%%%%%%%%%%%%%%%%%%%%%%%%%%%%%%%%%%%%%%%%%%%%%%%%%%%%%%%%%%%%%%%%%%%%%%%%%%%%%%%%%%%%%%%%%%%%%%%%%%%%%%

%\pagebreak

\noindent {\bf 6.7 Proof of theorem \ref{th:MDI_GEV}}

We show that the integral $K_n^-$, giving the contribution to the normalising constant from $\xi < -1$, diverges.
From the proof of theorem \ref{th:general_GEV} we have
\beqnn
K_n^- &=& 
(n-1)! \int_{-\infty}^{-1} \e^{-\gamma (1+\xi)} \,(-\xi)^{1-n}  \int_{y_n}^{\infty} \left\{\prod_{i=1}^n |y_i-\phi|^{-(1+1/\xi)}\right\} 
\left\{ \sum_{i=1}^n |y_i-\phi|^{-1/\xi} \right\}^{\!\!-n}
{\rm ~d}\phi {\rm ~d}\xi.
\eeqnn
For $\xi < -1$ we have $-(1+1/\xi)<0$ and $-1/\xi>0$.  Therefore, for $i=2, \ldots, n$,
$(\phi-y_i)^{-(1+1/\xi)} > (\phi-y_1)^{-(1+1/\xi)}$
and
$(\phi-y_i)^{-1/\xi} < (\phi-y_1)^{-1/\xi}$, and thus the
$\phi$-integrand is greater than $n^{-n} (\phi-y_1)^{-n}$.
Therefore,
\beqnn
K_n^- &>& 
(n-1)! \int_{-\infty}^{-1} \e^{-\gamma (1+\xi)} \,(-\xi)^{1-n}  \int_{y_n}^{\infty}
n^{-n} (\phi-y_1)^{-n} {\rm ~d}\phi {\rm ~d}\xi, \\
&=& (n-1)! \, n^{-n} (n-1)^{-1} (y_n-y_1)^{1-n} \int_{-\infty}^{-1} \e^{-\gamma(1+\xi)} \, (-\xi)^{1-n} {\rm ~d}\xi, \\
&=& (n-2)! \, n^{-n} (y_n-y_1)^{1-n} \e^{-\gamma} \int_{1}^{\infty} x^{1-n} \, \e^{\gamma x}  {\rm ~d}x,
\eeqnn
where $x=-\xi$.  For all samples sizes $n$ this integral diverges so the result follows.
\qed

%%%%%%%%%%%%%%%%%%%%%%%%%%%%%%%%%%%%%%%%%%%%%%%%%%%%%%%%%%%%%%%%%%%%%%%%%%%%%%%%%%%%%%%%%%%%%%%%%%%%%%%%%%%%%%%%%%%%%%%%%%%%

\noindent {\bf 6.8 Proof of theorem \ref{th:F_GEV}}

We need to show that $K_4$ is finite.
We split the range of integration over $\xi$ in (\ref{eqn:GEVa}) so that
$K_4=J_1+J_2+J_3$, with respective contributions from $\xi <-1$, $-1 \leq \xi \leq 0$ and $\xi >0$.

\noindent {\bf Proof that $J_1$ is finite.}  
We use the substitution $u=(\phi-y_1)^{-1}$ in (\ref{eqn:GEVa}) to give
\beqnn
J_1 
&=& 
3! \int_{-\infty}^{-1}  (-\xi)^{-3}  \int_{y_4}^{\infty}  \left\{\prod_{i=1}^4 (\phi-y_i)^{-(1+1/\xi)}\right\} 
\left\{ \sum_{i=1}^4 (\phi-y_i)^{-1/\xi} \right\}^{\!\!-n}
{\rm ~d}\phi {\rm ~d}\xi, \\
&=& 
3! \int_{-\infty}^{-1}  (-\xi)^{-3} \int_0^{1/\delta_4}
u^2 \prod_{i=2}^4 (1-\delta_i u)^{-(1+1/\xi)} 
\left\{ 1+\sum_{i=2}^4 (1-\delta_i u)^{-1/\xi} \right\}^{\!\!-4}
{\rm ~d}u {\rm ~d}\xi.  
\eeqnn
A similar calculation to (\ref{eqn:mine}) gives
\[ \prod_{i=2}^4 (1-\delta_i u)^{-(1+1/\xi)} \leq u^{-2(1+1/\xi)} 
\left\{\prod_{i=2}^{3}(\delta_4-\delta_i)\right\}^{\!\!-(1+1/\xi)} 
(1-\delta_4 u)^{-(1+1/\xi)}. \]
Noting also that $1+\sum_{i=2}^4 (1-\delta_i u)^{-1/\xi}\geq1$ we have
\beqnn
J_1 
&\leq&
3! \int_{-\infty}^{-1}  (-\xi)^{-3}
\left\{\prod_{i=2}^{3}(\delta_4-\delta_i)\right\}^{\!\!-(1+1/\xi)}
\int_0^{1/\delta_4}
u^{-2/\xi} (1-\delta_4 u)^{-(1+1/\xi)}
{\rm ~d}u {\rm ~d}\xi, \\
&=&   
3! \int_{-\infty}^{-1}  (-\xi)^{-3}
\left\{\prod_{i=2}^{3}(\delta_4-\delta_i)\right\}^{\!\!-(1+1/\xi)}
\beta \int_0^{1/\delta_4}
u^{-2/\xi} \frac{1}{\beta} \left(1+\frac{\xi u}{\beta}\right)^{-(1+1/\xi)}
{\rm ~d}u {\rm ~d}\xi, \\
&=& 
3! \int_{-\infty}^{-1}  (-\xi)^{-3}
\left\{\prod_{i=2}^{3}(\delta_4-\delta_i)\right\}^{\!\!-(1+1/\xi)}
\delta_n^{2/\xi-1} \frac{\Gamma(1-2/\xi) \Gamma(-1/\xi)}{\Gamma(1-3/\xi)}
{\rm ~d}\xi, 
\eeqnn
where $\beta=-\xi/\delta_4$ and the last line follows from (\ref{eqn:GPmom}) with $a=2$ and $\sigma=\beta$.

Therefore,
\beqn
J_1 &\leq&
3! \int_{-\infty}^{-1}  (-\xi)^{-3}
(y_4-y_1)^{2/\xi-1} \prod_{i=2}^3 (y_4-y_i)^{-(1+1/\xi)}
%(y_4-y_2)^{-(1+1/\xi)}(y_4-y_3)^{-(1+1/\xi)}
\frac{\Gamma(1-2/\xi) \Gamma(-1/\xi)}{\Gamma(1-3/\xi)}
{\rm ~d}\xi, \nonumber \\
&=& 
3! \prod_{i=1}^3 (y_4-y_i)^{-1} \int_{-\infty}^{-1}  (-\xi)^{-3} 
\left(\prod_{i=2}^3 \frac{y_4-y_i}{y_4-y_1}\right)^{\!\!\!-1/\xi}
\frac{\Gamma(1-2/\xi) \Gamma(-1/\xi)}{\Gamma(1-3/\xi)}
{\rm ~d}\xi, \nonumber \\
&=& 
3! \prod_{i=1}^3 (y_4-y_i)^{-1} \int_{0}^{1}  x
\left(\prod_{i=2}^3 \frac{y_4-y_i}{y_4-y_1}\right)^{\!\!\!x}
\frac{\Gamma(1+2 x) \Gamma(x)}{\Gamma(1+3 x)}
{\rm ~d}x, \nonumber \\
&=& 
3! \prod_{i=1}^3 (y_4-y_i)^{-1} \int_{0}^{1}  
\left(\prod_{i=2}^3 \frac{y_4-y_i}{y_4-y_1}\right)^{\!\!\!x}
\frac{\Gamma(1+2 x) \Gamma(1+x)}{\Gamma(1+3 x)}
{\rm ~d}x, \label{eqn:finiteGEV}
\eeqn
where $x=-1/\xi$ and we have used the relation $\Gamma(1+x)=x\,\Gamma(x)$.
The integrand in (\ref{eqn:finiteGEV}) is finite over the range of integration so this integral is finite and therefore $J_1$ is finite. 

\noindent {\bf Proof that $J_2$ is finite.}
Using the substitution $u=(\phi-y_1)^{-1}$ in (\ref{eqn:GEVa}) gives
\beqnn
J_2 
&=& 
3! \int_{-1}^{0}  (-\xi)^{-3} \int_0^{1/\delta_4}
u^2 \prod_{i=2}^4 (1-\delta_i u)^{-(1+1/\xi)} 
\left\{ 1+\sum_{i=2}^4 (1-\delta_i u)^{-1/\xi} \right\}^{\!\!-4}
{\rm ~d}u {\rm ~d}\xi.  
\eeqnn
For $-1 \leq \xi \leq 0$ we have $-(1+1/\xi) \geq 0$.  Noting that $0 < 1-\delta_i u < 1$ gives
\[ \prod_{i=2}^4 (1-\delta_i u)^{-(1+1/\xi)}  \leq (1-\delta_4 u)^{-(1+1/\xi)}. \]
Noting also that $1+\sum_{i=2}^4 (1-\delta_i u)^{-1/\xi}\geq1$ we have
\beqnn
J_2 &\leq&
3! \int_{-1}^{0}  (-\xi)^{-3}
\int_0^{1/\delta_4} u^2 (1-\delta_4u)^{-(1+1/\xi)} {\rm ~d}u {\rm ~d}\xi, \\
&=& 
3! \int_{-1}^{0}  (-\xi)^{-3}
\beta \int_0^{1/\delta_4} u^2 \frac{1}{\beta} \left(1+\frac{\xi u}{\beta}\right)^{-(1+1/\xi)} {\rm ~d}u {\rm ~d}\xi, \\
&=& 
3! \delta_4^{-3} \int_{-1}^{0} \frac{2}{(1-\xi)(1-2\xi)}  {\rm ~d}\xi, \\
&=& 12 (y_4-y_1)^{-3} \ln(3/2)
\eeqnn
where $\beta=-\xi/\delta_4$ and the penultimate line follows from (\ref{eqn:GPmom}) with $r=2$ and $\sigma=\beta$.

\noindent {\bf Proof that $J_3$ is finite.}
Using the substitution $u=(y_1-\phi)^{-1}$ in (\ref{eqn:GEVa}) gives
\beqnn
J_3
&=& 
3! \int_{0}^{\infty}  \xi^{-3}  \int_{-\infty}^{y_1}  \left\{\prod_{i=1}^4 (y_i-\phi)^{-(1+1/\xi)}\right\} 
\left\{ \sum_{i=1}^4 (y_i-\phi)^{-1/\xi} \right\}^{\!\!-4}
{\rm ~d}\phi {\rm ~d}\xi, \\
&=& 
3! \int_{0}^{\infty}  \xi^{-3} \int_0^{\infty}
u^2 \prod_{i=2}^4 (1+\delta_i u)^{-(1+1/\xi)} 
\left\{ 1+\sum_{i=2}^4 (1+\delta_i u)^{-1/\xi} \right\}^{\!\!-4}
{\rm ~d}u {\rm ~d}\xi.  
\eeqnn
Noting that for $\xi>0$ we have $-(1+1/\xi)<0$, using (\ref{eqn:mit}) with $a_k=\delta_k u$ gives
\[ \prod_{i=2}^4 (1+\delta_i u)^{-(1+1/\xi)} \leq (1+g u)^{-3(1+1/\xi)}, \]
where $g=(\delta_2\delta_3\delta_4)^{1/3}$.
Noting also that $1+\sum_{i=2}^4 (1+\delta_i u)^{-1/\xi}\geq1$ we have
\beqnn
J_3
&\leq&
3! \int_{0}^{\infty}  \xi^{-3} \int_0^{\infty}
u^2 (1+g u)^{-3(1+1/\xi)} 
{\rm ~d}u {\rm ~d}\xi, \\
&\leq&
3! \int_{0}^{\infty}  \xi^{-3} \beta \int_0^{\infty}
u^2 \frac{1}{\beta} \left(1+\frac{\alpha u}{\beta}\right)^{-(1+1/\alpha)} 
{\rm ~d}u {\rm ~d}\xi, 
\eeqnn
where $\alpha=\xi/(2\xi+3)$ and $\beta=\alpha/g$.
Therefore, (\ref{eqn:GPmomr}) with $r=2$, $\sigma=\beta$ and $\xi=\alpha$ gives
\beqnn
J_3 &\leq& 3!
\int_{0}^{\infty} \xi^{-3} \beta  \frac{2 \beta^2}{(1-\alpha)(1-2\alpha)} {\rm ~d}\xi, \\
&=&
4 g^{-3} \int_{0}^{\infty} \frac{1}{(\xi+3)(2\xi+3)} {\rm ~d}\xi, \\
&=&
\frac{4}{3} g^{-3} \int_{0}^{\infty} \left( \frac{1}{\xi+3/2} - \frac{1}{\xi+3} \right) {\rm ~d}\xi, \\
&=&
\frac{4}{3} g^{-3} \ln 2.
\eeqnn
The normalizing constant $K_4$ is finite, so $\pi_{U,GEV}(\mu,\sigma,\xi)$ yields a proper posterior density for $n=4$ and therefore does so for $n \geq 4$.
\qed

\noindent{\large\bf References}
\begin{description}
\bibitem[\protect\citeauthoryear{Abramowitz and Stegun}{Abramowitz and
  Stegun}{1972}]{AS1972}
Abramowitz, M. and I.~A. Stegun (1972).
\newblock {\em Handbook of Mathematical Functions with Formulas, Graphs, and
  Mathematical Tables}.
\newblock Dover Publications.

\bibitem[\protect\citeauthoryear{Alzer}{Alzer}{1999}]{Alzer1999}
Alzer, H. (1999).
\newblock Inequalities for the gamma function.
\newblock {\em Proc. Amer. Math. Soc.\/}~{\em 128\/}(1), 141--147.

\bibitem[\protect\citeauthoryear{Bayarri and Berger}{Bayarri and
  Berger}{2004}]{Bayarri2004}
Bayarri, M.~J. and J.~O. Berger (2004).
\newblock The interplay of {B}ayesian and frequentist analysis.
\newblock {\em Statist. Sci.\/}~{\em 19\/}(1), 58--80.

\bibitem[\protect\citeauthoryear{Beirlant, Goegebeur, Teugels, Segers, {De
  Waal}, and Ferro}{Beirlant et~al.}{2004}]{Beirlant2004}
Beirlant, J., Y.~Goegebeur, J.~Teugels, J.~Segers, D.~{De Waal}, and C.~Ferro
  (2004).
\newblock {\em {Statistics of Extremes}}.
\newblock John Wiley \& Sons.

\bibitem[\protect\citeauthoryear{Berger}{Berger}{2006}]{Berger2006}
Berger, J. (2006).
\newblock The case for objective {B}ayesian analysis.
\newblock {\em Bayesian analysis\/}~{\em 1\/}(3), 385--402.

\bibitem[\protect\citeauthoryear{Berger, Bernardo, and Sun}{Berger
  et~al.}{2009}]{BBS2009}
Berger, J., J.~Bernardo, and D.~Sun (2009).
\newblock The formal definition of reference priors.
\newblock {\em Ann. Statist.\/}~{\em 37}, 905--938.

\bibitem[\protect\citeauthoryear{Coles}{Coles}{2001}]{Coles2001}
Coles, S.~G. (2001).
\newblock {\em An Introduction to statistical modelling of extreme values}.
\newblock London: Springer.

\bibitem[\protect\citeauthoryear{Coles and Powell}{Coles and
  Powell}{1996}]{CP1996}
Coles, S.~G. and E.~A. Powell (1996).
\newblock Bayesian methods in extreme value modelling: a review and new
  developments.
\newblock {\em Int. Statist. Rev.\/}~{\em 64}, 119--136.

\bibitem[\protect\citeauthoryear{Coles and Tawn}{Coles and
  Tawn}{1996}]{Coles1996}
Coles, S.~G. and J.~A. Tawn (1996).
\newblock A {B}ayesian analysis of extreme rainfall data.
\newblock {\em J. R. Statist. Soc. C\/}~{\em 45\/}(4), 463--478.

\bibitem[\protect\citeauthoryear{Coles and Tawn}{Coles and Tawn}{2005}]{CT2005}
Coles, S.~G. and J.~A. Tawn (2005).
\newblock Bayesian modelling of extreme surges on the {UK} east coast.
\newblock {\em Phil. Trans. R. Soc. A.\/}~{\em 363\/}(1831), 1387--1406.

\bibitem[\protect\citeauthoryear{Datta, Mukerjee, Ghosh, and Sweeting}{Datta
  et~al.}{2009}]{DMGS2000}
Datta, G., R.~Mukerjee, M.~Ghosh, and T.~Sweeting (2009).
\newblock Bayesian prediction with approximate frequentist validity.
\newblock {\em Ann. Statist.\/}~{\em 28}, 1414--1426.

\bibitem[\protect\citeauthoryear{{Eugenia Castellanos} and Cabras}{{Eugenia
  Castellanos} and Cabras}{2007}]{EugeniaCastellanos2007}
{Eugenia Castellanos}, M. and S.~Cabras (2007).
\newblock {A default Bayesian procedure for the generalized Pareto
  distribution}.
\newblock {\em J. Statist. Plan. Infer.\/}~{\em 137\/}(2),
  473--483.

\bibitem[\protect\citeauthoryear{Fisher and Tippett}{Fisher and
  Tippett}{1928}]{FT1928}
Fisher, R.~A. and L.~H.~C. Tippett (1928).
\newblock Limiting forms of the frequency distribution of the largest or
  smallest member of a sample.
\newblock {\em Proc. Camb. Phil. Soc.\/}~{\em 24}, 180--190.

\bibitem[\protect\citeauthoryear{Giles and Feng}{Giles and Feng}{2009}]{GF2009}
Giles, D.~E. and H.~Feng (2009).
\newblock Bias-corrected maximum likelihood estimation of the parameters of the
  generalized {P}areto distribution.
\newblock Econometric Working Paper EWP0902, University of Victoria.

\bibitem[\protect\citeauthoryear{Gradshteyn and Ryzhik}{Gradshteyn and
  Ryzhik}{2007}]{GR1965}
Gradshteyn, I.~S. and I.~W. Ryzhik (2007).
\newblock {\em Table of Integrals, Series and Products\/} (7th ed.).
\newblock New York: Academic Press.

\bibitem[\protect\citeauthoryear{Ho}{Ho}{2010}]{Ho2010}
Ho, K.-W. (2010).
\newblock {A matching prior for extreme quantile estimation of the generalized
  Pareto distribution}.
\newblock {\em J. Statist. Plan. Infer.\/}~{\em 140\/}(6),
  1513--1518.

\bibitem[\protect\citeauthoryear{Hobert and Casella}{Hobert and
  Casella}{1996}]{Hobert1996}
Hobert, J.~P. and G.~Casella (1996).
\newblock The effect of improper priors on {G}ibbs sampling in hierarchical
  linear mixed models.
\newblock {\em J. Am. Statist.l Assoc.\/}~{\em
  91\/}(436), 1461--1473.

\bibitem[\protect\citeauthoryear{Hosking and Wallis}{Hosking and
  Wallis}{1987}]{Hosking1987}
Hosking, J. R.~M. and J.~R. Wallis (1987).
\newblock Parameter and quantile estimation for the generalized {P}areto
  distribution.
\newblock {\em Technometrics\/}~{\em 29\/}(3), 339--349.

\bibitem[\protect\citeauthoryear{Hosking, Wallis, and Wood}{Hosking
  et~al.}{1985}]{Hosking1985}
Hosking, J. R.~M., J.~R. Wallis, and E.~F. Wood (1985).
\newblock Estimation of the generalized extreme-value distribution by the
  method of probability-weighted moments.
\newblock {\em Technometrics\/}~{\em 27\/}(3), 251--261.

\bibitem[\protect\citeauthoryear{Jeffreys}{Jeffreys}{1961}]{Jeffreys1961}
Jeffreys, H. (1961).
\newblock {\em {Theory of Probability}}, Volume~2 of {\em The International
  series of monographs on physics}.
\newblock Oxford University Press.

\bibitem[\protect\citeauthoryear{Jenkinson}{Jenkinson}{1955}]{Jenkinson1955}
Jenkinson, A.~F. (1955).
\newblock The frequency distribution of the annual maximum (or minimum) values
  of meteorological elements.
\newblock {\em Q. J. R. Meteorol. Soc.\/}~{\em 81}, 158--171.

\bibitem[\protect\citeauthoryear{Kass and Wasserman}{Kass and
  Wasserman}{1996}]{Kass1996}
Kass, R.~E. and L.~Wasserman (1996).
\newblock The selection of prior distributions by formal rules.
\newblock {\em J. Am. Statist. Assoc.\/}~{\em
  91\/}(435), 1343--1370.

\bibitem[\protect\citeauthoryear{Kotz and Nadarajah}{Kotz and
  Nadarajah}{2000}]{KN2000}
Kotz, S. and S.~Nadarajah (2000).
\newblock {\em Extreme value distributions: theory and applications}.
\newblock London: Imperial College Press.

\bibitem[\protect\citeauthoryear{Leadbetter, Lindgren, and
  Rootz\'{e}n}{Leadbetter et~al.}{1983}]{LLR1983}
Leadbetter, M.~R., G.~Lindgren, and H.~Rootz\'{e}n (1983).
\newblock {\em Extremes and Related Properties of Random Sequences and
  Processes}.
\newblock New York: Springer.

\bibitem[\protect\citeauthoryear{Martins and Stedinger}{Martins and
  Stedinger}{2000}]{Martins2000}
Martins, E.~S. and J.~R. Stedinger (2000).
\newblock {Generalized maximum-likelihood generalized extreme-value quantile
  estimators for hydrologic data}.
\newblock {\em Water Resour. Res.\/}~{\em 36\/}(3), 737.

\bibitem[\protect\citeauthoryear{Martins and Stedinger}{Martins and
  Stedinger}{2001}]{Martins2001}
Martins, E.~S. and J.~R. Stedinger (2001).
\newblock Generalized maximum likelihood {P}areto-{P}oisson estimators for
  partial duration series.
\newblock {\em Water Resour. Res.\/}~{\em 37\/}(10), 2551.

\bibitem[\protect\citeauthoryear{Mitrinovi\'{c}}{Mitrinovi\'{c}}{1964}]{Mitrinovic1964}
Mitrinovi\'{c}, D.~A. (1964).
\newblock {\em Elementary inequalities}.
\newblock Groningen: Noordhoff.

\bibitem[\protect\citeauthoryear{Pickands}{Pickands}{1975}]{Pickands1975}
Pickands, J. (1975).
\newblock Statistical inference using extreme order statistics.
\newblock {\em Ann. Statist.\/}~{\em 3}, 119--131.

\bibitem[\protect\citeauthoryear{Pickands}{Pickands}{1994}]{Pickands1994}
Pickands, J. (1994).
\newblock Bayes quantile estimation and threshold selection for the generalized
  {P}areto family.
\newblock In J.~Galambos, J.~Lechner, and E.~Simiu (Eds.), {\em Extreme Value
  Theory and Applications}, pp.\  123--138. Springer US.

\bibitem[\protect\citeauthoryear{Qiu and Vuorinen}{Qiu and
  Vuorinen}{2004}]{QV2004}
Qiu, S.~L. and M.~Vuorinen (2004).
\newblock Some properties of the gamma and psi functions, with applications.
\newblock {\em Math. Comp.\/}~{\em 74\/}(250), 723--742.

\bibitem[\protect\citeauthoryear{Roy and Dey}{Roy and Dey}{2014}]{RD2014}
Roy, V. and D.~K. Dey (2014).
\newblock Propriety of posterior distributions arising in categorical and
  survival models under generalized extreme value distribution.
\newblock {\em Statist. Sinica\/}~{\em 24}, 699--722.

\bibitem[\protect\citeauthoryear{Smith}{Smith}{2005}]{Smith2005}
Smith, E. (2005).
\newblock {\em Bayesian Modelling of Extreme Rainfall Data}.
\newblock Ph.\ D. thesis, University of Newcastle upon Tyne.

\bibitem[\protect\citeauthoryear{Smith}{Smith}{1984}]{Smith1984}
Smith, R.~L. (1984).
\newblock Threshold methods for sample extremes.
\newblock In J.~Oliveira (Ed.), {\em Statistical Extremes and Applications},
  Volume 131 of {\em NATO ASI Series}, pp.\  621--638. Springer Netherlands.

\bibitem[\protect\citeauthoryear{Smith}{Smith}{1985}]{Smith1985}
Smith, R.~L. (1985).
\newblock Maximum likelihood estimation in a class of non-regular cases.
\newblock {\em Biometrika\/}~{\em 72}, 67--92.

\bibitem[\protect\citeauthoryear{Smith}{Smith}{1989}]{Smith1989}
Smith, R.~L. (1989).
\newblock Extreme value analysis of environmental time series: An application
  to trend detection in ground-level ozone.
\newblock {\em Statist. Sci.\/}~{\em 4}, 367--377.

\bibitem[\protect\citeauthoryear{Smith and Goodman}{Smith and
  Goodman}{2000}]{SG2000}
Smith, R.~L. and D.~J. Goodman (2000).
\newblock Bayesian risk analysis.
\newblock In P.~Embrechts (Ed.), {\em Extremes and Integrated Risk Management},
  pp.\  235--251. London: Risk Books.

\bibitem[\protect\citeauthoryear{Stephenson and Tawn}{Stephenson and
  Tawn}{2004}]{ST2004}
Stephenson, A. and J.~A. Tawn (2004).
\newblock Inference for extremes: accounting for the three extremal types.
\newblock {\em Extremes\/}~{\em 7\/}(4), 291--307.

\bibitem[\protect\citeauthoryear{Tuyl, Gerlachy, and Mengersenz}{Tuyl
  et~al.}{2009}]{TGM2009}
Tuyl, F., R.~Gerlachy, and K.~Mengersenz (2009).
\newblock Posterior predictive arguments in favor of the {B}ayes-{L}aplace
  prior as the consensus prior for binomial and multinomial parameters.
\newblock {\em Bayesian analysis\/}~{\em 4\/}(1), 151--158.

\bibitem[\protect\citeauthoryear{van Noortwijk, Kalk, and Chbab}{van Noortwijk
  et~al.}{2004}]{van2004}
van Noortwijk, J.~M., H.~J. Kalk, and E.~H. Chbab (2004).
\newblock Bayesian estimation of design loads.
\newblock {\em HERON\/}~{\em 49\/}(2), 189--205.

\bibitem[\protect\citeauthoryear{Yang and Berger}{Yang and
  Berger}{1998}]{YB1998}
Yang, R. and J.~Berger (1998).
\newblock A catalog of noninformative priors.
\newblock ISDS Discussion paper 97-42, Duke University.\\
\newblock Available at
  http://www.stats.org.uk/priors/noninformative/YangBerger1998.pdf.

\bibitem[\protect\citeauthoryear{Zellner}{Zellner}{1971}]{Zellner1971}
Zellner, A. (1971).
\newblock {\em {An Introduction to Bayesian Inference in Econometrics}},
  Volume~17 of {\em Wiley Series in Probability and Mathematical Statistics}.
\newblock John Wiley and Sons.

\bibitem[\protect\citeauthoryear{Zellner}{Zellner}{1998}]{Zellner1998}
Zellner, A. (1998).
\newblock Past and recent results on maximal data information priors.
\newblock {\em J. Statist. Res.\/}~{\em 32\/}(1), 1--22.
\end{description}

\end{document}